\documentclass[12pt]{article}
\usepackage[nosort]{cite}
\usepackage{pstricks}
\usepackage{graphicx}
\usepackage{amsmath}
\usepackage{amssymb}

\newcommand{\be}{\begin{equation}}
\newcommand{\ee}{\end{equation}}
\newcommand{\bea}{\begin{eqnarray}}
\newcommand{\eea}{\end{eqnarray}}

\begin{document}

\setcounter{page}{0}
\thispagestyle{empty}
%\pagestyle{plain}
%\pagestyle{empty}

%%%%%%%%%%%%%%%%%%%%%%%%%%%%%%%%%%%%%%%%%%%%%%%%%%%%%%%%%%%%%%%%%%%%%%%%%%%%%%%
\begin{flushright}
ULB-TH/14-04\;
USM-TH-320
\end{flushright}
\vskip 8pt

\begin{center}
{\bf \LARGE {Radiative neutrino mass generation linked to  \\ \vskip 8pt
 neutrino mixing and $0\nu\beta\beta$-decay predictions}}
\end{center}

\vskip 12pt

\begin{center}
{\bf   Michael Gustafsson$^{a}$ Jose M. No$^{b}$ and Maximiliano A. Rivera$^{c}$}
\end{center}

\vskip 20pt

\begin{center}

\centerline{$^{a}${\it Service de Physique Th\'eorique, Universit\'e Libre de Bruxelles, B-1050 Bruxelles, Belgium}}
\centerline{$^{b}${\it Department of Physics and Astronomy, University of Sussex, BN1 9QH Brighton, UK}}
\centerline{$^{c}${\it Departamento de F\'isica, Universidad T\'ecnica Federico Santa Mar\'ia}}
\centerline{{\it Casilla 110-V, Valparaiso, Chile}}

\vskip .3cm
\centerline{\tt mgustafs@ulb.ac.be, J.M.No@sussex.ac.uk, maximiliano.rivera@usm.cl}
\end{center}

\vskip 13pt

\begin{abstract}
We discuss the connection between the origin of neutrino masses and their mixings which arises in a class of scenarios with radiatively induced neutrino masses. In these scenarios, the neutrino mass matrix acquires textures with two entries close to zero in the basis where the charged-lepton mass matrix is diagonal. This results in specific constraints on the neutrino mixing parameters, which leads to the prediction of (i) a normal ordering of neutrino masses with the lightest neutrino mass in the $\sim$\;meV range and (ii) testable correlations among the various mixing angles, including a nonzero  $\theta_{13}$ angle with its exact value correlated with the values of the atmospheric angle $\theta_{23}$ and the $CP$ phase $\delta$. We quantify the impact of deviations from exact zeroes in the mass matrix texture and connect it to the amount of hierarchy among Yukawa couplings. These scenarios of radiative neutrino mass generation also give rise to new short-range contributions to neutrinoless double beta 
decay, which dominate over the usual light-neutrino exchange contribution. As a result, this class of models can have a sizable neutrinoless double beta decay rate, in the range of upcoming experiments despite the normal mass ordering of neutrinos.
\end{abstract}

\newpage
\renewcommand{\theequation}{\arabic{section}.\arabic{equation}}

%%%%%%%%%%%%%%%%%%%%%%%%%%%%%%%%%%%%%%%%%%%%%%%%%%%%%%%%%%%%
\section{Introduction} 
%%%%%%%%%%%%%%%%%%%%%%%%%%%%%%%%%%%%%%%%%%%%%%%%%%%%%%%%%%%%
A central aspect of neutrino physics is the study of the origin and structure of the neutrino mass matrix and  its connection to the Dirac or Majorana nature of the neutrinos.  
Several appealing mechanisms have been advocated to account for the tiny but nonzero masses of neutrinos, ranging from the seesaw mechanism (see Ref.\ \cite{Mohapatra:2005wg} for a review) to radiative neutrino mass generation (see, {\it e.g.}, Refs.\ \cite{Zee:1980ai,Babu:1988ki,Krauss:2002px,Ma:2006km,Aoki:2008av}). The vast majority of these scenarios predicts that neutrinos are of Majorana nature, but observationally it remains an open question. At the same time, the improved precision in the observed pattern of neutrino mixings has triggered a lot of theoretical activity aimed at explaining its origin based on ideas such as flavor symmetries (see Ref.\ \cite{Morisi:2012fg} for a recent overview) or even anarchy \cite{Hall:1999sn}. 

It is not inconceivable that the same underlying new physics could be responsible for both the $10^{-2}$ -- $10^{-1}$ eV neutrino mass scale and the observed pattern of neutrino masses and mixings, which would then provide a unified understanding of both aspects of neutrino physics. The purpose of this work is to explore a simple class of scenarios beyond the Standard Model (SM) in which this connection between the origin of neutrino masses and neutrino mixing naturally arises in the context of radiative neutrino mass generation. As we shall see, these scenarios also incorporate a particular link to the neutrinoless double beta ($0\nu\beta\beta$) decay probe of the Majorana nature of neutrinos (as discussed in Sec.~\ref{sec:0nu2beta}).

As a starting point in our analysis, we consider how specific textures of the neutrino mass matrix $m^\nu$ can be responsible for the neutrino mass scale and mixing patterns.  We focus on scenarios in which the neutrino mass matrix has one or several entries that are either exactly or approximate zero, so called ``texture zeroes'' \cite{Fritzsch:1977za,Fritzsch:1977vd,Weinberg:1977hb,Wilczek:1977uh,Frampton:2002yf}.
%, Guo:2002ei, Merle:2006du,Dev:2006qe,Fritzsch:2011qv,Ludl:2011vv,Grimus:2012zm,Grimus:2012ii}
Specifically, we look at cases with $m^{\nu}_{ee} \simeq 0$ and $m^{\nu}_{e\mu} \simeq 0$ \cite{Frampton:2002yf,Xing:2002ta,Xing:2002ap,Desai:2002sz,Guo:2002ei,Honda:2003pg,Merle:2006du,Dev:2006xu,Dev:2006qe,Fritzsch:2011qv,Kumar:2011vf,Ludl:2011vv,Meloni:2012sx, Grimus:2012zm}
(the precise definition of $m^\nu$ and an overview of its observational constraints from neutrino oscillation data will be presented in Sec.~\ref{sec:data}).
Those scenarios induce nontrivial correlations among the various neutrino oscillation parameters, and can accommodate all the current observational data while giving interesting testable predictions for some of the unknowns in the neutrino sector.  Of special interest are the predicted neutrino mass ordering and the correlation between the values of the reactor angle $\theta_{13}$, the octant of the atmospheric angle $\theta_{23}$, and the Charge-Parity ($CP$) phase $\delta$.

It is then shown that such neutrino mass textures are obtained in a class of models of radiative neutrino mass generation. Concrete model examples \cite{Chen:2006vn,delAguila:2011gr,delAguila:2012nu,Gustafsson:2012vj} of such class thus provide attractive examples of embedding  neutrino mixing properties into a theory of neutrino mass generation. In these setups, neutrino masses are generated at either 2- or 3-loop order, providing an elegant explanation for the smallness of neutrino masses compared to the electroweak scale \cite{Farzan:2012ev}. %(some of these scenarios also incorporate an additional link to dark  matter particle candidates \cite{Gustafsson:2012vj}).       

Finally, a generic feature of these scenarios is the existence of an important contribution to $0\nu\beta\beta$ decay from short-distance physics effects \cite{Chen:2006vn,delAguila:2011gr,delAguila:2012nu}, resulting in potentially large amplitudes for $0\nu\beta\beta$ decay processes despite the fact that the standard contribution from light-neutrino exchange is extremely suppressed (since $m^{\nu}_{ee} \simeq 0$). This fact could make these processes detectable in ongoing and upcoming $0\nu\beta\beta$ decay experiments, including GERDA\cite{Schonert:2005zn,Agostini:2013mzu}, EXO\cite{Akimov:2005mq}, SNO+\cite{Chen:2005yi}, KamLAND-Zen\cite{Mitsui:2011zza,KamLANDZen:2012aa}, CUORE\cite{Arnaboldi:2002du}, NEXT\cite{Dafni:2012nwa,GomezCadenas:2012jv}, MAJORANA \cite{Gaitskell:2003zr}, and SuperNEMO\cite{Arnold:2010tu}. Remarkably, such scenarios give a potentially detectable signal in $0\nu\beta\beta$ decay experiments together with a normal ordering of the neutrino masses and a small lightest 
neutrino mass $m_1 < 10^{-2}$ eV.

\bigskip
The paper is organized as follows: In Sec.~\ref{sec:data}, we introduce our conventions for the neutrino mass matrix and review the present experimental situation for neutrino masses and mixings from neutrino oscillations.  In Sec.~\ref{sec:textures}, the correlations among different neutrino parameters are studied in detail for scenarios in which $m^{\nu}_{ee} = 0$, with emphasis on the scenario $m^{\nu}_{ee} = m^{\nu}_{e\mu} = 0$. In Sec.~\ref{sec:rad_text} we identify and explore a class of radiative neutrino mass generation scenarios that naturally generate a neutrino mass matrix with approximate texture zeroes of the same form as those studied in Sec.~\ref{sec:textures}. We then define a technical measure of the amount of hierarchy (in the neutrino ``Yukawa'' matrix) for these scenarios and show that it is indeed milder than in the charged-lepton sector. In Sec.~\ref{sec:0nu2beta} we explore the features of the leading, short-distance contribution to $0\nu\beta\beta$ decay in these scenarios and derive  
prospects of detection in various present and future $0\nu\beta\beta$ decay experiments. Finally, we conclude in Sec.~\ref{sec:conclusion}.

%%%%%%%%%%%%%%%%%%%%%%%%%%%%%%%%%%%%%%%%%%%%%%%%%%%%%%%%%%%%
\section{Conventions and neutrino oscillation data }\label{sec:data}
%%%%%%%%%%%%%%%%%%%%%%%%%%%%%%%%%%%%%%%%%%%%%%%%%%%%%%%%%%%%
For the case of Majorana neutrinos, a parametrization of their mass matrix, in the basis where charged current interactions are flavor diagonal and the charged leptons $e, \mu, \tau$ are simultaneously mass eigenstates, reads
\be
\label{UPMNS}
m^{\nu} = U^{T} \,m^{\nu}_{D} \,U \quad \mathrm{with} \quad m^{\nu}_{D} = \mathrm{Diag}\left(m_1,m_2,m_3\right).
\ee
Here $m_{1,2,3}$ are the masses of the three light neutrinos, and $U^{T}$ is the Pontecorvo-Maki-Nakagawa-Sakata (PMNS) matrix \cite{Pontecorvo:1957qd}, given in terms of three mixing angles $\theta_{12}$, $\theta_{23}$, $\theta_{13}$ and three phases (a $CP$ phase $\delta$ and two Majorana phases\footnote[1]{We adopt here the convention for the Majorana phases given in Ref.~\cite{Merle:2006du}.} $\alpha_1$ and $\alpha_2$),
\bea
\label{UPMNS2}
U = \mathrm{Diag}\left(1,e^{i\alpha_{1}},e^{i(\alpha_{2}+\delta)}\right) \times \quad \quad \quad 
\quad \quad \quad \nonumber\\ 
\left( \begin{array}{ccc}
c_{13}c_{12} & -c_{23}s_{12}- s_{23}c_{12}s_{13}e^{i\delta} & s_{23}s_{12}- c_{23}c_{12}s_{13}e^{i\delta} \\    
c_{13}s_{12}  & c_{23}c_{12}- s_{23}s_{12}s_{13}e^{i\delta} &  -s_{23}c_{12}- c_{23}s_{12}s_{13}e^{i\delta} \\
s_{13}e^{-i\delta} & s_{23}c_{13}  & c_{23}c_{13}
        \end{array}
 \right),
\eea
with $s_{ij} \equiv \mathrm{sin}(\theta_{ij})$ and $c_{ij} \equiv \mathrm{cos}(\theta_{ij})$. Investigating the presence (or absence) of an organizing principle behind the observed structure in $m^{\nu}_{D}$ and the PMNS matrix $U$ is then a central aspect of neutrino phenomenology. This in turn requires an accurate experimental determination of the various neutrino parameters, in particular the neutrino mass ordering, the three mixing angles, and the $CP$ phase.

Until the year 2011, there existed only an experimental upper bound on the value of the mixing angle $\theta_{13}$, while $\theta_{23}$ and $\theta_{12}$ were relatively well determined and consistent with the tribimaximal \cite{Harrison:2002er} neutrino mixing hypothesis $\theta_{12} = 30^{\circ}$,  $\theta_{23} = 45^{\circ}$, $\theta_{13} = 0^{\circ}$. However, the recent experimental data from Daya-Bay \cite{An:2012eh}, RENO \cite{Ahn:2012nd}, Double Chooz \cite{Abe:2011fz}, T2K \cite{Abe:2011sj}, and MINOS \cite{Adamson:2011qu}, measuring a nonzero value for $\theta_{13}$, combined with the latest results from atmospheric neutrinos \cite{Wendell:2010md,Itow:2013zza}, which are possibly suggesting a departure of the atmospheric angle  $\theta_{23}$ from its maximal mixing value $\pi/4$, have provided us with a new perspective on neutrino mixing. This is summarized in the up-to-date global fits to neutrino oscillation data in Refs.~\cite{Tortola:2012te,Fogli:2012ua,GonzalezGarcia:2012sz}.
Of them,  we shall use Ref.~\cite{GonzalezGarcia:2012sz}, which gives $\Delta m^2_{21} \equiv m^2_{2} - m^2_{1} = 7.50^{+0.18}_{-0.19}\times 10^{-5} \mathrm{eV}^2$, $\left|\Delta m^2_{31}\right| \equiv \left| m^2_{3} - m^2_{1} \right|= 2.473^{+0.07}_{-0.067}\times 10^{-3} \mathrm{eV}^2$  ($2.427^{+0.065}_{-0.042}\times 10^{-3} \mathrm{eV}^2$) for $\Delta m^2_{31} > 0$ ($\Delta m^2_{31}<0$),  $s_{12}^2 = 0.302^{+0.013}_{-0.012}$, $s_{13}^2 = 0.0227^{+0.0023}_{-0.0024}$, and $s_{23}^2 = 0.413^{+0.037}_{-0.025}$ ($0.594^{+0.021}_{-0.022}$) if on the first (second) octant for $\theta_{23}$. 

Neutrino oscillation experiments are still not sensitive to the sign of $\Delta m^2_{31}$ which results in two possible mass orderings in the neutrino sector, commonly known as normal ordering (NO) and inverted ordering (IO), that are characterized by
\be
\begin{array}{l}
\Delta m^2_{31} > 0 \quad \rightarrow \quad  m_1 < m_2 < m_3 \quad \mathrm{(NO)}\\
\Delta m^2_{31} < 0 \quad \rightarrow \quad   m_3 < m_1 < m_2 \quad \mathrm{(IO)\,.}
\end{array}
\ee
Also, while nonmaximal $\theta_{23}$ mixing seems now favored, current experimental data are not sensitive to the sign of its deviation from maximality (see \cite{Tortola:2012te,GonzalezGarcia:2012sz} for a discussion on this issue). Finally, the value of the $CP$ phase $\delta$ is also beyond current experimental sensitivity, although it is expected that future measurements from T2K and NO$\nu$A will begin to constrain the $CP$ phase.

%%%%%%%%%%%%%%%%%%%%%%%%%%%%%%%%%%%%%%%%%%%%%%%%%%%%%%%%%%%%
\section{Zeroes in the neutrino mass matrix} \label{sec:textures}
%%%%%%%%%%%%%%%%%%%%%%%%%%%%%%%%%%%%%%%%%%%%%%%%%%%%%%%%%%%%

We will now discuss the phenomenological aspects of models which have certain entries $m^{\nu}_{ab}$ in the neutrino mass matrix of Eq.~(\ref{UPMNS}) exactly or approximately equal to zero. We present an up-to-date review analysis of textures with $m^{\nu}_{ee} \simeq 0$ and especially when jointly with $m^{\nu}_{e\mu} \simeq 0$ (these textures have been widely studied in the literature; see, 
{\it e.g.}, Refs.~\cite{Frampton:2002yf,Xing:2002ta,Xing:2002ap,Desai:2002sz,Guo:2002ei,Honda:2003pg,Merle:2006du,Dev:2006qe,Fritzsch:2011qv,Kumar:2011vf,Ludl:2011vv,Meloni:2012sx, Grimus:2012zm,Grimus:2012ii} 
and references therein). These textures are relevant for the class of models we will study in the following sections. There are only three textures with $m^{\nu}_{ee} = 0$ allowed by neutrino oscillation data \cite{Frampton:2002yf,Grimus:2012zm}: 

%We now motivate and discuss scenarios in which some of the entries of the neutrino mass matrix (\ref{UPMNS}) are exactly or approximately zero. We will analyze mass textures for $m^{\nu}_{ab}$ with $m^{\nu}_{ee} \simeq 0$. There are only three such textures allowed by neutrino oscillation data \cite{Frampton:2002yf}: 
%
\be
\label{Textures}
\left( \begin{array}{ccc}
0 & \times & \times \\    
\times & \times & \times \\
\times & \times & \times
 \end{array}
 \right), \quad \quad 
\left( \begin{array}{ccc}
0 & 0 & \times \\    
0 & \times & \times \\
\times & \times & \times
 \end{array}
 \right), \quad \quad 
\left( \begin{array}{ccc}
0 & \times & 0 \\    
\times & \times & \times\\
0 & \times & \times
 \end{array}
 \right),  
\ee
where the $\times$ denote nonvanishing (not necessarily equal) entries.  In this section we focus on the case of exact texture zeroes and leave the discussion of approximate zeroes and its connection to radiative generation of neutrino masses for Sec.~\ref{sec:rad_text}. The (complex) neutrino mass matrix is, for the case of Majorana neutrinos, a function of nine independent parameters,  of which six are measurable via neutrino oscillations.\footnote[2]{Neutrino oscillations are not sensitive to the Majorana phases $\alpha_1$ and $\alpha_2$ nor to the absolute neutrino mass scale.}  Setting any matrix element to zero then gives two equations (both its real and imaginary part have to vanish), which impose correlations among various neutrino parameters.  From Eq.~(\ref{UPMNS2}), the textures in (\ref{Textures}) give the following relations: 
\\
\be
\label{zeroee}
\hspace{-7.0cm}
m^{\nu}_{ee} \equiv c_{13}^2 \left(m_1 c_{12}^2 + e^{2i \alpha_1}  m_2  s_{12}^2 \right) + 
 e^{2i \alpha_2}  m_3  s_{13}^2 = 0\,, 
\ee
\\
\vspace{-7mm}
\be
\label{zeroemu}
m^{\nu}_{e\mu} \equiv c_{13} \left[ \left(e^{2i \alpha_1} m_2 - m_1 \right) s_{12} c_{12} c_{23} 
+ e^{i \delta} s_{23} s_{13} \left( e^{2 i \alpha_2}  m_3 -  
m_1 c_{12}^2 - e^{2i \alpha_1} m_2  s_{12}^2\right) \right] = 0 \, ,
\ee
\\
\vspace{-7mm}
\be
\label{zeroetau}
m^{\nu}_{e\tau} \equiv c_{13} \left[ \left(m_1 - e^{2i \alpha_1} m_2 \right) s_{12} c_{12} s_{23} 
+ e^{i \delta} c_{23} s_{13} \left( e^{2 i \alpha_2}  m_3 -  
m_1 c_{12}^2 - e^{2i \alpha_1} m_2  s_{12}^2\right) \right] = 0 \, .
\ee

\medskip
%An immediate question that arises is if the above textures give a prediction for the neutrino mass ordering. 
From the condition (\ref{zeroee}), which is common to all three textures, we already obtain information on possible ranges of neutrino masses and mixing angles. In particular, from Eq.~(\ref{zeroee}) it follows that $\theta_{13}$ is constrained to 
\be
\label{throat1}
\frac{\left|s_{12}^2\,m_2 - c_{12}^2\,m_1 \right|}{m_3} 
\leq t^2_{13} \leq 
\frac{s_{12}^2\,m_2 + c_{12}^2\,m_1}{m_3} 
\ee
with $t_{13} \equiv \mathrm{tan} (\theta_{13})$. This allows us to obtain in a 
straightforward manner the allowed range of solutions for normal (NO) and inverted (IO) ordering in the ($m,\theta_{13}$) plane, where $m$ is the mass of the lightest neutrino:
\begin{itemize}
\item NO: $m_1 = m$,\; $m_2 = \sqrt{\Delta m^2_{21} + m^2}$,\; and $m_3 = \sqrt{\Delta m^2_{31} + m^2}\,$;
\item IO:\space{}  $m_1 = \sqrt{\left|\Delta m^2_{31}\right| + m^2}$,\;  $m_2 = \sqrt{\left|\Delta m^2_{31}\right| + \Delta m^2_{21} + m^2}$,\; and $m_3 = m\,$.
\end{itemize}

\begin{figure}[t]
\center{\includegraphics[width=.7 \columnwidth]{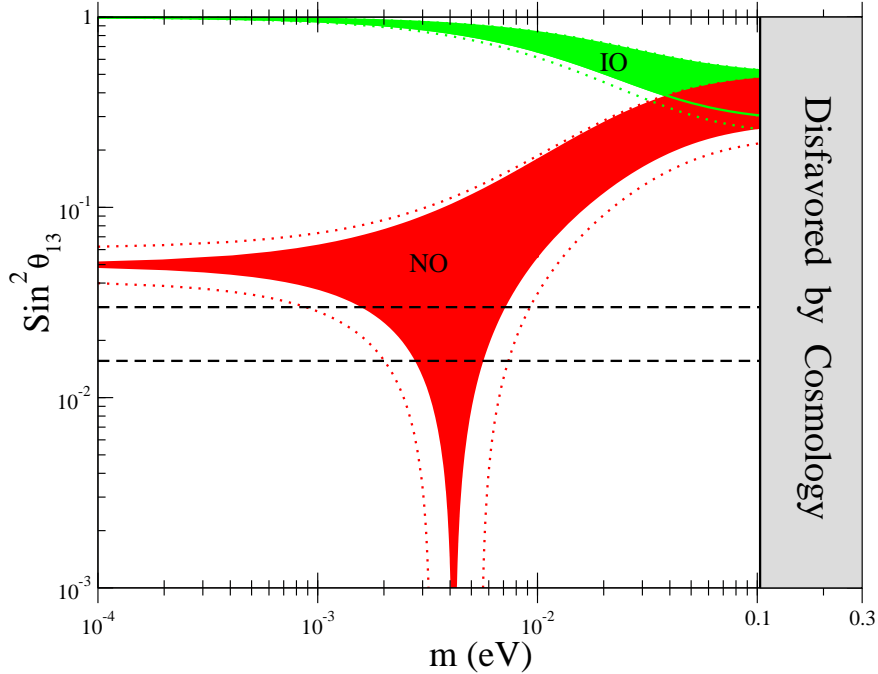}}
\caption{\small Allowed regions for $\theta_{13}$ as a function of the lightest neutrino mass $m$ for a neutrino mass matrix with texture $m_{ee}^\nu =0$. The regions are for normal (NO/red) and inverted (IO/green) mass ordering, with best-fit measured values of $\theta_{12}$, $\Delta m^2_{13}$, and $\Delta m^2_{21}$ (solid region) as given in  Ref.~\cite{GonzalezGarcia:2012sz}. The allowed region with $\theta_{12}$, $\Delta m^2_{13}$, and $\Delta m^2_{21}$ within their $3\sigma$  experimental ranges (dotted line) is also shown. The dashed-black horizontal lines enclose the $3\sigma$ allowed experimental range for $\theta_{13}$. The grey  region is disfavored by cosmological data \cite{Ade:2013zuv}.}
\label{fig1:ee}
\end{figure}

The range of solutions to Eq.~(\ref{throat1}) for NO and IO are shown in Fig.~\ref{fig1:ee}, both for $\theta_{12}$, $\Delta m^2_{31}$ and $\Delta m^2_{21}$ set to their best-fit values (solid regions) and with these parameters allowed to vary independently within their $3\sigma$ experimental ranges (inside dotted lines).
%%%%%%%%%%%%%%%% 
%%%%%%%%%%%%%%%%
As can be seen from Fig~\ref{fig1:ee}, the experimental constraint on the reactor mixing angle $\theta_{13}$ excludes the  inverted mass ordering scenario when $m^{\nu}_{ee} = 0$, and thus all the three textures in (\ref{Textures}) predict a normal ordering for the neutrino masses. From Fig.~\ref{fig1:ee} it is also clear that Eq.~(\ref{zeroee}) is only satisfied in a limited range of values for $m$ as a function of $\theta_{13}$ \cite{Merle:2006du,Pascoli:2001by}  (this range is commonly  referred to as ``the chimney"). For the oscillation parameters within their $3\sigma$ allowed experimental  range, $m$ must lie in the interval $0.001\;\mathrm{eV} \lesssim m \lesssim 0.009\; \mathrm{eV}$.

%\medskip 
As can be seen from Eq.~(\ref{zeroee}), $m^{\nu}_{ee}$ does not depend on the atmospheric mixing angle $\theta_{23}$ or the $CP$ phase $\delta$. We will now analyze how an extra zero in the texture leads to further relations among the various neutrino oscillation parameters, which also involve $\theta_{23}$ and $\delta$.

%%%%%%%%%%%%%%%%%%%%%%%%%%%%%%%%%%%%%%%%%%%%%%%%%%%%%%%%%%%%
\subsection{Mass matrix texture with $m^{\nu}_{ee} = 0$ and  $m^{\nu}_{e\mu} = 0$}
%%%%%%%%%%%%%%%%%%%%%%%%%%%%%%%%%%%%%%%%%%%%%%%%%%%%%%%%%%%%

For the case of two independent texture zeroes in $m^{\nu}$, such as Eqs.~(\ref{zeroee}) and (\ref{zeroemu}), these  form a system of four equations (vanishing real and imaginary parts of $m^{\nu}_{ee}$, and $m^{\nu}_{e\mu}$). This system of equations links the values of  $\delta$, $\alpha_1$, $\alpha_2$ and $m$ to the values of the neutrino oscillation parameters $\Delta m^2_{21}$, $\Delta m^2_{31}$, $\theta_{12}$,  $\theta_{23}$, and $\theta_{13}$ and leads to specific predictions of $m$, $\delta$,  $\alpha_1$, and $\alpha_2$ (up to an overall sign change in $\delta$, $\alpha_1$ and $\alpha_2$) when the other parameters are specified.  Moreover, solutions will only exist for certain values of $\Delta m^2_{21}$, $\Delta m^2_{31}$, $\theta_{12}$, $\theta_{23}$ and $\theta_{13}$, so the imposition of such textures results in nontrivial relations among these parameters.\footnote[3]{Neutrino mass matrix textures with more than two entries set to zero would overconstrain the system and have no solution for $m_1$,
 $\delta$, $\alpha_1$, and $\alpha_2$ given the current neutrino oscillation data \cite{Frampton:2002yf,Grimus:2012zm}.}

\begin{figure}[t]
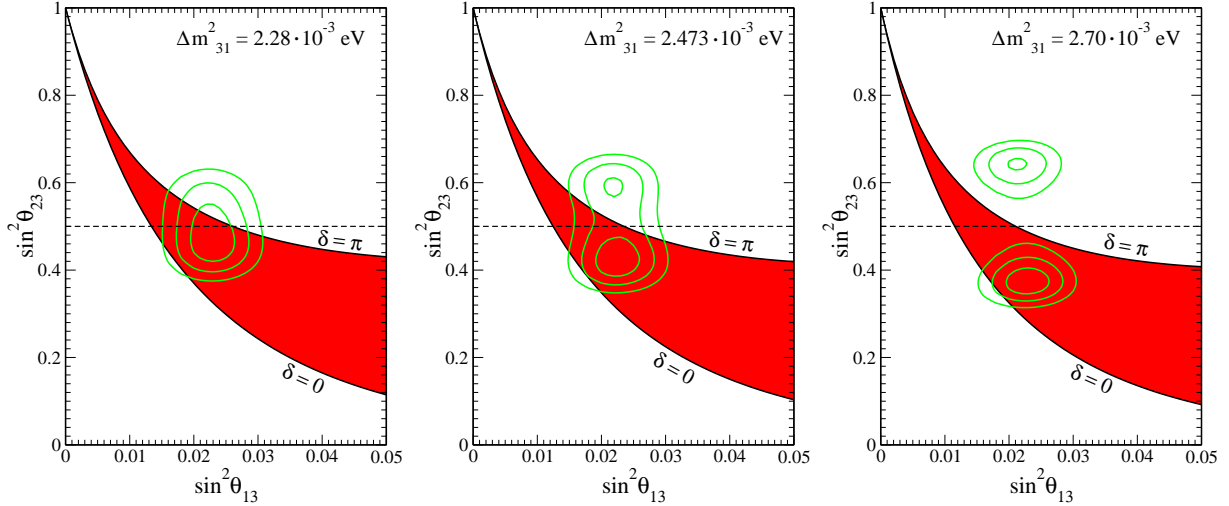

\includegraphics[width=.32 \columnwidth]{figs/Memu_2.eps}
\includegraphics[width=.32 \columnwidth]{figs/Memu_1.eps}
\includegraphics[width=.32 \columnwidth]{figs/Memu_3.eps}
\caption{\small Allowed regions for the neutrino mass texture $m^{\nu}_{ee} = m^{\nu}_{e\mu} = 0$ in the ($s^2_{13}$, $s^2_{23}$) plane (solid-red), for $\Delta m^2_{31}$ fixed at its $-3\sigma$ value (LEFT), best-fit value (MIDDLE) and $+3\sigma$ value (RIGHT), and best-fit values of $\Delta m^2_{21}$ and $s^2_{12}$. In each case, the $1\sigma$, $2\sigma$ and $3\sigma$ allowed experimental contours from the global analysis of Ref.~\cite{GonzalezGarcia:2012sz} are shown in green. The horizontal dashed line indicates maximal mixing $\theta_{23} = \pi/4$.}
\label{fig1:texture}
\end{figure}

\begin{figure}[t]
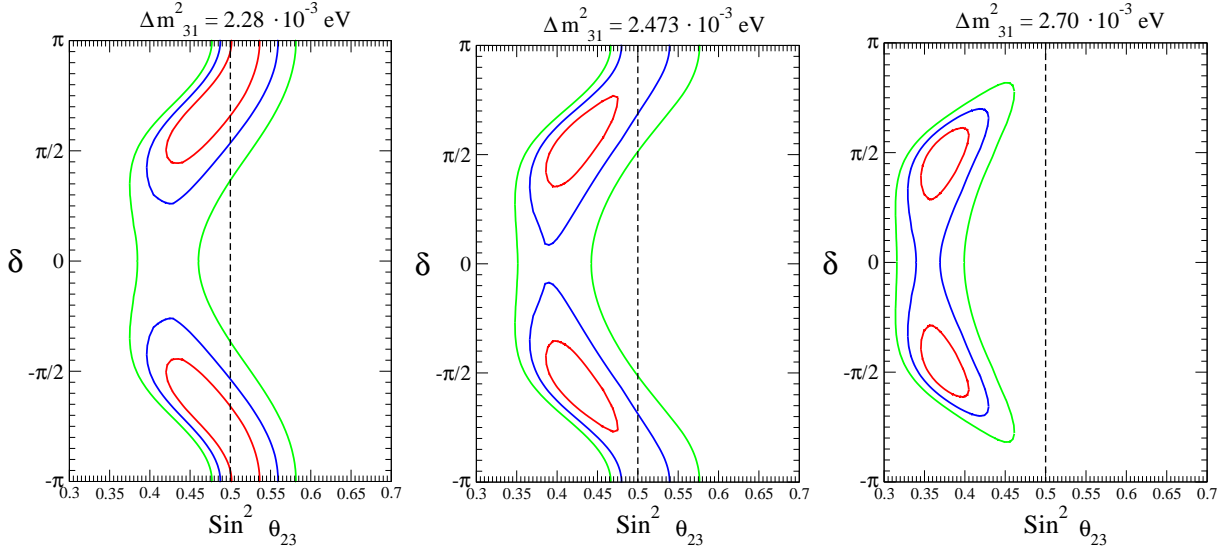

\includegraphics[width=.32 \columnwidth]{figs/23_delta_A.eps}
\includegraphics[width=.32 \columnwidth]{figs/23_delta_B.eps}
\includegraphics[width=.32 \columnwidth]{figs/23_delta_C.eps}
\caption{\small Allowed values of the $CP$ phase $\delta$ (for the neutrino mass 
texture $m^{\nu}_{ee} = m^{\nu}_{e\mu} = 0$) as a function of $s^2_{23}$ along the $1\sigma$ (red), $2\sigma$ (blue), and $3\sigma$ (green) experimental contours from Fig.~\ref{fig1:texture}. 
The values of $\Delta m^2_{31}$, $\Delta m^2_{21}$, and $s^2_{12}$ are set as in Fig.~\ref{fig1:texture}. The vertical dashed line indicates maximal mixing $\theta_{23} = \pi/4$.} 
\label{fig2:texture}
\end{figure}

As is clear from the discussion in the previous section, the requirement $m^{\nu}_{ee} = 0$ resulted in specific relations among $\theta_{12}$, $\theta_{13}$, and the lightest neutrino mass $m$. Requiring also $m^{\nu}_{e\mu} = 0$ results in additional correlations, involving now also the atmospheric mixing angle $\theta_{23}$, the $CP$ phase $\delta$, and the other neutrino oscillation parameters. In particular, further requiring $m^{\nu}_{e\mu} = 0$ correlates the allowed values of $\theta_{13}$, $\theta_{23}$, and $\delta$, as shown in Figs.~\ref{fig1:texture} and  \ref{fig2:texture}. In addition, the allowed range for the value of the lightest neutrino mass scale $m$ in Fig.~\ref{fig1:ee} gets slightly narrowed down once the experimental bounds on $\theta_{23}$ are imposed. 
%%%%%%%%%%%%%%%% 
%%%%%%%%%%%%%%%%
%%%%%%%%%%%%%%%% 
%%%%%%%%%%%%%%%%

From Eqs. (\ref{zeroee}) and (\ref{zeroemu}), it is possible to derive a ``master formula" correlating all the neutrino oscillation parameters (see Appendix~\ref{sec:app1} for details). For a normal ordering,
\be
\label{master_1}
\frac{\Delta m^2_{21}}{\Delta m^2_{31}} = s_{13}^2\,\frac{t_{23}^2}{t_{12}^2}\, 
\frac{1 - t_{12}^4 + 2 s_{13}\, t_{12}^2 \, t_{23}^{-1} \left(t_{12}^{-1} + t_{12} \right) \cos \delta } 
{c_{13}^4 -s_{13}^2 \left( t_{12}^2 \,t_{23}^2 - 2s_{13} \,t_{12} \, t_{23} \cos \delta +s_{13}^2\right)}\, .
\ee
It is important to stress that the relation (\ref{master_1}) does not include any assumption about the values of the elements of the neutrino mass matrix $m^{\nu}_{ab}$ other than 
$m^{\nu}_{ee}$, $m^{\nu}_{e\mu}$. We emphasize that this relation does not predict specific values for the neutrino oscillation parameters. Nevertheless, 
Eq. (\ref{master_1}) predicts strong correlations among the different oscillation parameters which might not have been compatible with current neutrino oscillation experimental data (see Appendix A).
Instead, it successfully accommodates the neutrino oscillation data and gives clear predictions for the correlations among future precision data in the neutrino sector. For small values of $\theta_{13}$, the right-hand side of Eq.~(\ref{master_1}) can be expanded in powers of $s_{13}$, which to lowest orders then reads
\be
\label{master_2}
\frac{\Delta m^2_{21}}{\Delta m^2_{31}} = s_{13}^2 \,\frac{t_{23}^2}{t_{12}^2}\,
\left[\left(1 - t_{12}^4 \right) + 2\, s_{13}\, \frac{t_{12}\, \left(1 + t_{12}^2 \right)}{t_{23}}\, 
 \cos \delta \right] + \mathcal{O}(s_{13}^4)\,.
\ee

In Fig.~\ref{fig1:texture} we show the allowed region (solid-red) from Eq.~(\ref{master_1}) in the ($s^2_{13}$, $s^2_{23}$) plane  for $\Delta m^2_{31}$ fixed at its best-fit value (MIDDLE) and at its -/+$3\sigma$ value (LEFT/RIGHT). Both $\Delta m^2_{21}$ and $s^2_{12}$ are fixed to their best-fit values in Fig.~\ref{fig1:texture} (but a marginalization over these parameters would also give almost identical experimental contour regions). The upper bound on $s^2_{23}$ as a function  of $s^2_{13}$ corresponds to $\delta = \pm \pi$, and the lower bound corresponds to $\delta = 0$, which  can be understood from Eq.~(\ref{master_2}). Also shown are the $1\sigma$, $2\sigma$, and $3\sigma$  experimentally allowed regions (green lines) from the global analysis of \cite{GonzalezGarcia:2012sz,NUFIT}.

As can be seen from Fig.~\ref{fig1:texture}, the allowed range of the atmospheric mixing  angle $\theta_{23}$ for a neutrino mass texture with $m^{\nu}_{ee} = m^{\nu}_{e\mu} = 0$ is not symmetric around $\theta_{23} = \pi/4$  ($s^2_{23} = 1/2$), and a value of $\theta_{23} < \pi/4$ seems favored given current experimental  constraints on the value of $\theta_{13}$. This preference for the lower  $\theta_{23}$ octant is only very mild for $\Delta m^2_{31} = 2.28 \cdot 10^{-3}$ eV  (and in fact, in this case it is not very well motivated to talk  about the $\theta_{23}$ octant, since the preferred region for $\theta_{23}$ lies close to $\pi/4$) and becomes more pronounced for increasing $\Delta m^2_{31}$. 
The conclusion of a preferred $\theta_{23} < \pi/4$ depends also weakly on the precise value of the solar angle $\theta_{12}$, since for values of $\theta_{12}$ significantly larger than the  best-fit value the red region in Fig.~\ref{fig1:texture} is shifted upward and eventually covers a sizeable part of the upper octant as well. Thus, a future, more accurate determination of the value of $\theta_{12}$ could either reinforce this conclusions or make the preference for the lower $\theta_{23}$-octant in this texture eventually disappear.

The mass texture with $m^{\nu}_{ee} = m^{\nu}_{e\mu} = 0$ also results in interesting correlations among  the values of $\theta_{13}$, $\theta_{23}$, and the $CP$ phase $\delta$, as shown in Fig.~\ref{fig2:texture}. Here the $1\sigma$, $2\sigma$, and $3\sigma$ allowed experimental regions of $s^2_{13}$ {\it vs} $s^2_{23}$ are plotted in the ($\delta$, $s^2_{23}$) plane  for the same values of  $\Delta m^2_{31}$, $\Delta m^2_{21}$, $s^2_{12}$ as adopted in the three plots in Fig.~\ref{fig1:texture} [note that since for $m^{\nu}_{ee} = m^{\nu}_{e\mu} = 0$ there are only two independent parameters out of the three parameters $\theta_{13}$, $\theta_{23}$, $\delta $ we can always extract the prediction of $\delta$ along any contour in the ($s^2_{13}$, $s^2_{23}$) plane].
We see that values of $\left|\delta\right| \sim \pi/2$ become more favored for increasing $\Delta m^2_{31}$, while for $\Delta m^2_{31} \sim 2.28 \cdot 10^{-3}$ eV values $\left|\delta\right| > \pi/2$, and in particular $\left|\delta\right| \rightarrow \pi$, are preferred. Fig.~\ref{fig2:texture} also shows a specific correlation between the values of $\theta_{23}$ and $\delta$ within this texture: larger values of $\theta_{23}$ seem to favor larger values of $\left|\delta\right|$. Finally, it is important to  notice that the texture does not impose any restriction on the sign of the $CP$ phase, since $\delta$  appears in Eq.~(\ref{master_1}) as $\cos \delta$.  
%{\red These results update and refines previous similar studies, such as \cite{Dev:2006xu,Dev:2006qe} where {\it e.g.}\ $\theta_{13}$ was still unknown and the texture prediction was to have a nonzero reactor mixing angle.} 

%%%%%%%%%%%%%%%%%%%%%%%%%%%%%%%%%%%%%%%%%%%%%%%%%%%%%%%%%%%%
\subsection{Mass matrix texture with  $m^{\nu}_{ee} = 0$ and  $m^{\nu}_{e\tau} = 0$.}
%%%%%%%%%%%%%%%%%%%%%%%%%%%%%%%%%%%%%%%%%%%%%%%%%%%%%%%%%%%%

The entry $m^{\nu}_{e\tau}$ of the neutrino mass matrix is similar in structure to $m^{\nu}_{e\mu}$ and can in fact be obtained from the latter via the substitutions $\theta_{23} \rightarrow \pi/2 - \theta_{23}$ and  $\delta \rightarrow \delta + \pi$ \cite{Fritzsch:2011qv}. Then the correlation between $\theta_{13}$ and $\theta_{23}$ for $m^{\nu}_{ee} = m^{\nu}_{e\tau} = 0$ is reversed with respect to the one found for $m^{\nu}_{ee} = m^{\nu}_{e\mu} = 0$ in the previous section, since now Fig.~\ref{fig1:texture} and \ref{fig2:texture} would look the same but with the replacement $s^2_{23} \rightarrow c^2_{23}$.   In particular, for $m^{\nu}_{ee} = m^{\nu}_{e\tau} = 0$, $\theta_{23} > \pi/4$ (the second octant) is now favored. Furthermore, both entries $m^{\nu}_{e\tau}$ and $m^{\nu}_{e\mu}$ cannot vanish at the same time. For $m^{\nu}_{ee} = 0$ it can be shown that 
\begin{multline}
\label{emu_etau}
\quad \left|m^{\nu}_{e\mu}\right|^2 + \left|m^{\nu}_{e\tau}\right|^2 = \\  c^2_{13} \left[\left( s_{12} \, c_{12}
\right)^2 \left(\sqrt{\Delta m^2_{21} + m^2} - m\right)^2
%\right. 
%\left. 
+ s_{13}^2 \left(1+ t^2_{13} \right)^2 
\left(\Delta m^2_{31} + m^2\right)\right]. \quad
\end{multline}
For $\Delta m^2_{21}$, $\Delta m^2_{31}$, $s^2_{12}$, and $s^2_{13}$ inside their 3$\sigma$ allowed experimental ranges, we find the following bound $\sqrt{\left|m^{\nu}_{e\mu}\right|^2 + \left|m^{\nu}_{e\tau}\right|^2} \geq 0.0063$ eV, where the inequality is saturated for $m \lesssim 0.009$~eV.  Therefore, a texture with $m^{\nu}_{ee} = m^{\nu}_{e\mu}  = m^{\nu}_{e\tau} = 0$ is not possible.

\bigskip

\begin{figure}[t]
\center{\includegraphics[width=.85 \columnwidth]{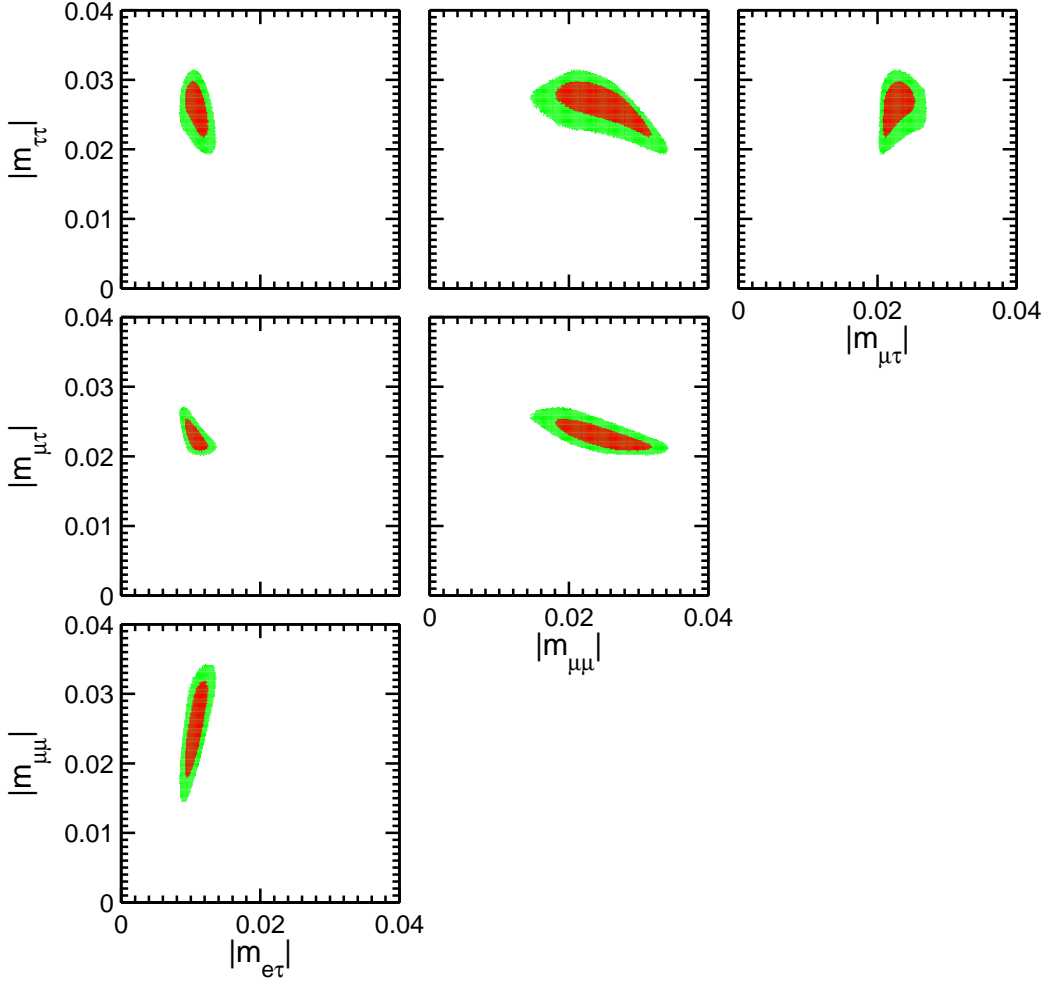}}
\caption{\small The size of the entries in the neutrino mass matrix $\left|m^{\nu}_{e\tau}\right|$, 
$\left|m^{\nu}_{\mu\mu}\right|$, $\left|m^{\nu}_{\mu\tau}\right|$ and $\left|m^{\nu}_{\tau\tau}\right|$ 
(in eV) for mass textures  with $m^{\nu}_{ee} = m^{\nu}_{e\mu} = 0$. The red regions correspond to the allowed values for neutrino oscillation
data within $1\sigma$ confidence level, while the green regions correspond to the allowed values for neutrino oscillation data within $3\sigma$ confidence level.} 
\label{fig3:texture}
\end{figure}

It is worth stressing that if $m^{\nu}_{ee} = 0$ and either of  $m^{\nu}_{e\mu} = 0$ or $m^{\nu}_{e\tau} = 0$ are imposed the allowed ranges for the rest of the  neutrino mass matrix entries are very constrained (once the current neutrino oscillation allowed experimental ranges are also imposed). This can be explicitly seen in Fig.~\ref{fig3:texture} for the case $m^{\nu}_{ee} = m^{\nu}_{e\mu} = 0$, where the allowed ranges of the neutrino mass matrix entries $\left|m^{\nu}_{e\tau}\right|$,  $\left|m^{\nu}_{\mu\mu}\right|$, $\left|m^{\nu}_{\mu\tau}\right|$, and $\left|m^{\nu}_{\tau\tau}\right|$ are computed from a scan over all experimentally allowed values of the neutrino oscillation parameters $\Delta m^2_{21}$, $\Delta m^2_{31}$, $s^2_{12}$, $s^2_{23}$ and $s^2_{13}$. 
%%%%%%%%%%%%%%%% 
%%%%%%%%%%%%%%%%
To perform the scan, random initial conditions are given to the unconstrained parameters $\delta$, $\alpha_{1}$, $\alpha_{2}$, and $m$ ($<$ 0.1 eV), and  we generate $\gtrsim 10^6$ numerical solutions to  $m^{\nu}_{ee} = m^{\nu}_{e\mu} = 0$. The 1$\sigma$ and 3$\sigma$ allowed regions for the neutrino mass matrix entries $|m_{ab}^\nu|$ are found from those random-scan points by calculating each scan point's total log likelihood relative to the best-fit value $\Delta \ln \mathcal{L}$ and finding the points with $-2 \Delta \ln \mathcal{L} < 5.88$ and 18.2, respectively. These limiting values on $2 \Delta \ln \mathcal{L}$ are the appropriate values for 5 degrees of freedom, corresponding to the five measured neutrino-oscillation parameters. Technically speaking, the total likelihood function $\mathcal{L}$ is built up from the product of uncorrelated single Gaussian probability distribution functions (pdfs) for each observable, except for the $\theta_{23}$ pdf which is instead modelled as the sum of two 
properly normalized Gaussian pdfs with minima in two separate octants and a third Gaussian pdf around the maximal mixing value $\pi/4$, all in order to properly match the result presented in \cite{NUFIT}.\footnote[4]{We use their ``Huber Fluxes, no Reactor-Short-Base-Line (RSBL)'' v1.2 results.}
A comparison to, {\it e.g.}, Refs.~\cite{Meloni:2012sx,Grimus:2012zm} shows that their $m^\nu$ matrices have entries $m^\nu_{ab}$ that fall well within the contour regions shown in Fig.~\ref{fig3:texture}.

%%%%%%%%%%%%%%%%%%%%%%%%%%%%%%%%%%%%%%%%%%%%%%%%%%%%%%%%%%%%
\section{Textures from radiative neutrino mass generation} \label{sec:rad_text}
%%%%%%%%%%%%%%%%%%%%%%%%%%%%%%%%%%%%%%%%%%%%%%%%%%%%%%%%%%%%

We now discuss the connection between the neutrino mass textures analyzed in the previous section and scenarios of neutrino mass generation. In particular, we will show that a certain type of scenario  for radiative neutrino mass generation leads to approximate mass textures of the form (\ref{Textures}). 

\begin{figure}[t]
\center{\includegraphics[width=.65 \columnwidth]{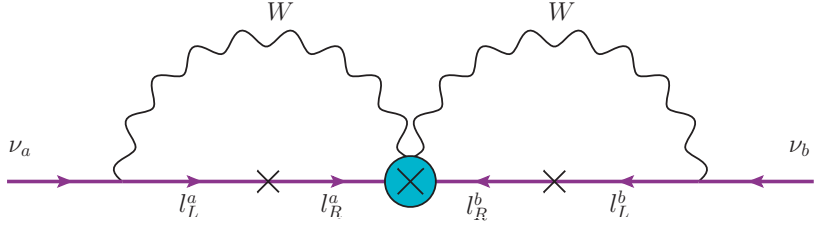}}
\caption{\small The 2-loop diagram generating neutrino masses from $\mathcal{O}^9$ in Eq.~(\ref{D9}). The crosses $\times$ in the fermion propagators represent chirality flips, which (upon electroweak symmetry breaking) make this Feynman amplitude proportional to the masses $m_{l_a}$ and $m_{l_b}$.} 
\label{fig:TwoLoops}
\end{figure}

Let us begin by simply noting that beyond SM (BSM) physics must be lepton number violating (LNV) in order to allow for the generation of Majorana masses for the light neutrinos. Then, by assuming that no extra gauge symmetries beyond the  electroweak $SU(2)_L \times U(1)_Y$ are present, we can parametrize the effect of the LNV new physics in terms of  nonrenormalizable operators that include only SM fields and preserve all the local symmetries of the SM.  We assume that the LNV physics couples only directly to leptons, but not directly to quarks. Under these generic assumptions, it was recently shown in Ref.~\cite{delAguila:2012nu} that when the LNV physics couples only directly to leptons of right-handed chirality $\ell_{R}$  (and not to those of left-handed chirality), then the only lowest-order operator (appearing at dimension $D = 9$)  that violates lepton number by two units ($\Delta L = 2$) is given by\footnote[5]{Note that this operator does not appear in the classification of SM $\Delta L = 2$ 
effective operators in Refs.~\cite{Babu:2001ex,Gouvea:2007xp}. LNV effective operators involving SM gauge bosons were thought to be unable to accommodate a suitable renormalizable completion in Ref.~\cite{Babu:2001ex} --see also Ref.~\cite{Angel:2012ug} for a brief discussion on this issue.}
\be
\label{D9}
\mathcal{O}^9 \equiv C^{(9)}_{ab} 
\,\overline{\ell^c}_{R_a} \ell_{R_b} \left[\left(D_{\mu} H\right)^T i\sigma_2 H \right]^2
\ee
with $C^{(9)}_{ab}$ being a matrix in flavor space. This is part of a more general result in Ref.~\cite{delAguila:2012nu} regarding the dimensions and structure of the lowest-order LNV non-renormalizable SM operators involving leptons and no quarks\footnote[6]{The lowest-order operator involving only leptons of right-handed chirality appears at $D = 9$, while the lowest-order operator involving both left- and right-handed chiralities appears at $D = 7$. The lowest-order operator involving only left-handed leptons is the well-known D=5 \textit{Weinberg operator}\cite{Weinberg:1979sa}.}. Upon electroweak symmetry breaking, $\mathcal{O}^9$ induces a term: 
\be
\label{D9WW}
\frac{C^{(9)}_{ab}}{\tilde{\Lambda}}\,\overline{\ell^c}_{R_a} \ell_{R_b} W^+_{\mu} W^{+\mu}.
\ee
We stress that 
the scale $\tilde{\Lambda}$ may not directly correspond to any specific new physics scale, but rather to a combination of different mass scales, and in particular it might be lower than any of those in scenarios with some hierarchy of scales. The term (\ref{D9WW}) generates a leading contribution to neutrino masses at 2 loops, with two chirality flips, as shown in Fig.~\ref{fig:TwoLoops}.
%%%%%%%%%%%%%%%% 
%%%%%%%%%%%%%%%%
Implications of $\mathcal{O}^9$ for $0\nu \beta\beta$ decay will be analyzed in Sec.~5. The neutrino mass matrix $m^{\nu}_{ab}$ is then proportional to the charged-lepton masses $m_{l_a}\, m_{l_b}$, since weak charged currents conserve lepton flavor,
\be
\label{loopnumass}
m^{\nu}_{ab} \sim \left(\frac{1}{16\, \pi^2}\right)^n \,C^{(9)}_{ab} \, \frac{m_{l_a}\, m_{l_b}}{\Lambda}  \, ,
\ee
where $\Lambda$ can be related to $\tilde{\Lambda}$. Both the loop suppression and $C^{(9)}_{ab}/\Lambda$ will ultimately depend\footnote[7]{While neutrino masses generated from $\mathcal{O}^9$ in Eq.~(\ref{D9}) appear at 2 loops, the operator $\mathcal{O}^9$ may itself have been generated at loop order in an underlying renormalizable completion (see Sec.~\ref{sec:BSM}), and thus $n \geq 2$.} on the specific LNV new physics responsible for generating $\mathcal{O}^9$. We note at this point that $m^{\nu}_{ab} \sim m_{l_a}\, m_{l_b}$ might also be achieved in other scenarios that do not involve  $\mathcal{O}^9$ (see {\it e.g.} the discussion in Ref.~\cite{Chang:1999hga}), but this is nevertheless a very nongeneric feature of radiative neutrino mass models.

\begin{figure}[t]
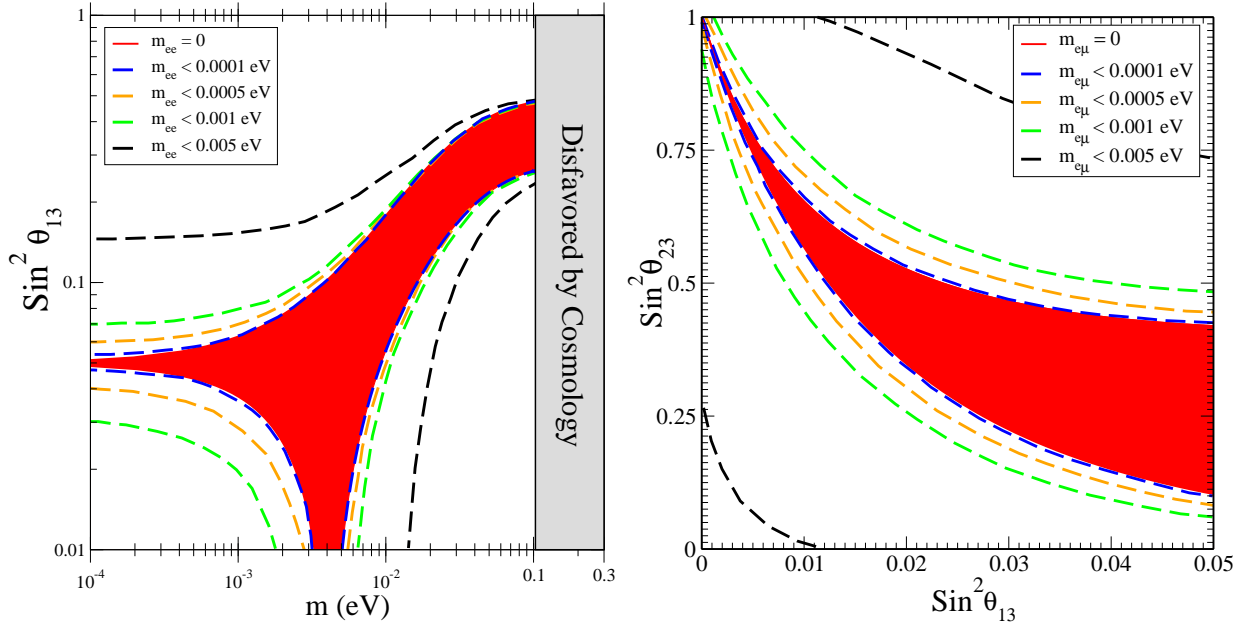

\includegraphics[width=.49 \columnwidth]{figs/m_eeF3_Nozero.eps}
\includegraphics[width=.49 \columnwidth]{figs/Fig_Memu_Nonzero.eps}

\caption{\small LEFT: For normal ordering (NO), the allowed region in the ($m, s^2_{13}$) plane as the upper bound on $\left|m^{\nu}_{ee}\right|$ is increased. RIGHT: For $m^{\nu}_{ee} = 0$, the allowed region in the ($s^2_{13},  \,s^2_{23}$) plane as the upper bound on $\left|m^{\nu}_{e\mu}\right|$ is increased. In both cases, we consider the best-fit values for the other neutrino oscillation parameters \cite{GonzalezGarcia:2012sz}.} 
\label{memu}
\end{figure}

It is apparent from Eq.~(\ref{loopnumass}) that due to the $m_{l_a}\, m_{l_b}$ dependence, the entries in the neutrino mass matrix $m^{\nu}_{ab}$ proportional to $m_{l_e}$ will be expected to be much smaller than the rest (up to the size of $C^{(9)}_{ab}$; see the discussion below), since $m_{l_{\tau}} \gg m_{l_{\mu}} \gg m_{l_e}$. This suppression of certain mass matrix elements leads to a matrix texture including approximate zeroes (the zeroes will not be exact unless the corresponding $C^{(9)}_{ab} = 0$). Nevertheless, the predictions for the neutrino mass ordering (NO {\it vs} IO), absolute neutrino mass scale $m$, octant of $\theta_{23}$ and $CP$ phase $\delta$ presented in Sec.~\ref{sec:textures} are still exactly verified for small enough entries. Fig.~\ref{memu} (LEFT and RIGHT) shows that this occurs for $\left| m^{\nu}_{ee}\right|, \left| m^{\nu}_{e\mu}\right| \lesssim 10^{-4}$ eV.  
%%%%%%%%%%%%%%%% 
%%%%%%%%%%%%%%%%
However, as the size of the entries $m^{\nu}_{ee}, \, m^{\nu}_{e\mu}$ increases, there is a transition region in which these predictions get  fainter, eventually disappearing for $\left| m^{\nu}_{ee}\right|, \left| m^{\nu}_{e\mu}\right| \gtrsim 10^{-3}$ eV. 

The appearance of approximate zeroes in the neutrino mass matrix (\ref{loopnumass}) depends on the size of  $C^{(9)}_{ab}/\Lambda$. Since $\left|m^{\nu}_{ee}\right| \lesssim 10^{-4}$~eV is needed (Fig.~\ref{memu} LEFT), we find a lower limit $\Lambda_L$ on the required effective scale $\Lambda/C^{(9)}_{ee}$, given approximately by $\Lambda_L \sim 100$ GeV, $600$ MeV, and $4$ MeV for $n = 2$, $3$, and $4$, respectively. A Similar argument can be applied in order to have $\left|m^{\nu}_{e\mu}\right| \lesssim 10^{-4}$~eV (Fig.~\ref{memu} RIGHT). As will be discussed in Sec.~\ref{sec:BSM}, LNV new physics giving a renormalizable completion to $\mathcal{O}^9$ involve new states with nontrivial $SU(2)_L \times U(1)_Y$ quantum numbers, as well as new states with sizable couplings 
to leptons. These scenarios are then difficult to realize with an effective scale lower than $\Lambda_L$, without being in conflict with high energy collider data.\footnote[8]{Recall, however, that $\Lambda$ does not directly correspond to a scale of new physics in the sense of effective field theory.} 
Renormalizable completions of $\mathcal{O}^9$ then predict $\left|m^{\nu}_{ee}\right| < 10^{-4}$~eV and strongly favor $\left|m^{\nu}_{e\mu}\right| < 10^{-4}$~eV. In this sense, the textures analyzed in the Sec.~\ref{sec:textures} are naturally realized for this class of scenarios.
%, {\red and typically $\left|m^{\nu}_{e\mu}\right| \lesssim 10^{-4}$~eV. }
%while for $\left| m^{\nu}_{e\mu}\right| \lesssim 10^{-4}$~eV a mild hierarchy between $C^{(9)}_{e\mu}$ and $C^{(9)}_{e\tau}$ is needed (see below).

On the other hand, being compatible with current neutrino oscillation data poses lower bounds to the rest of the entries of $m^{\nu}_{ab}$. This is due to the fact that oscillation data allow for at most two independent texture zeroes \cite{Frampton:2002yf,Grimus:2012zm} and require that the remaining entries in $m^{\nu}_{ab}$ are all of a size between  $\sqrt{\Delta m^2_{21}}$ and $\sqrt{\Delta m^2_{31}}$ (see Fig.~\ref{fig3:texture}). The most stringent bound for our purposes is $\left|m^{\nu}_{e\tau}\right| \gtrsim 0.008$ eV (recall Fig.~\ref{fig3:texture}), which imposes the requirement $\Lambda/C^{(9)}_{e\tau} < \Lambda_H$, with $\Lambda_H \sim 5$ TeV, $30$ GeV, and $200$ MeV  for $n = 2, 3$, and $4$, respectively. Satisfying these lower bounds on the remaining $m^{\nu}_{ab}$ together with $\left|m^{\nu}_{e\mu}\right| \lesssim 10^{-4}$~eV requires a certain amount of hierarchy\footnote[9]{In fact, for $m_{ee},m_{e\mu} \simeq 0$ the required hierarchy among the different entries $C^{(9)}_{ab}$ in Eq.~(\
ref{loopnumass}) can be directly derived from Fig.~\ref{fig3:texture}.} among the elements of $C^{(9)}_{ab}$ (in particular, $C^{(9)}_{e\mu} \lesssim C^{(9)}_{e\tau}/5$ is required).  We stress that this does not have an impact on the generation of the texture in these scenarios but rather implies that some hierarchy of couplings is presumably required in any beyond the Standard Model realization of $\mathcal{O}^9$  to fit current oscillation data.

%%%%%%%%%%%%%%%%%%%%%%%%%%%%%%%%%%%%%%%%%%%%%%%%%%%%%%%%%%%%
\subsection{Approximate textures and a measure of hierarchy}
%%%%%%%%%%%%%%%%%%%%%%%%%%%%%%%%%%%%%%%%%%%%%%%%%%%%%%%%%%%%

\begin{figure}[t]
\center{
\includegraphics[width=.43 \columnwidth]{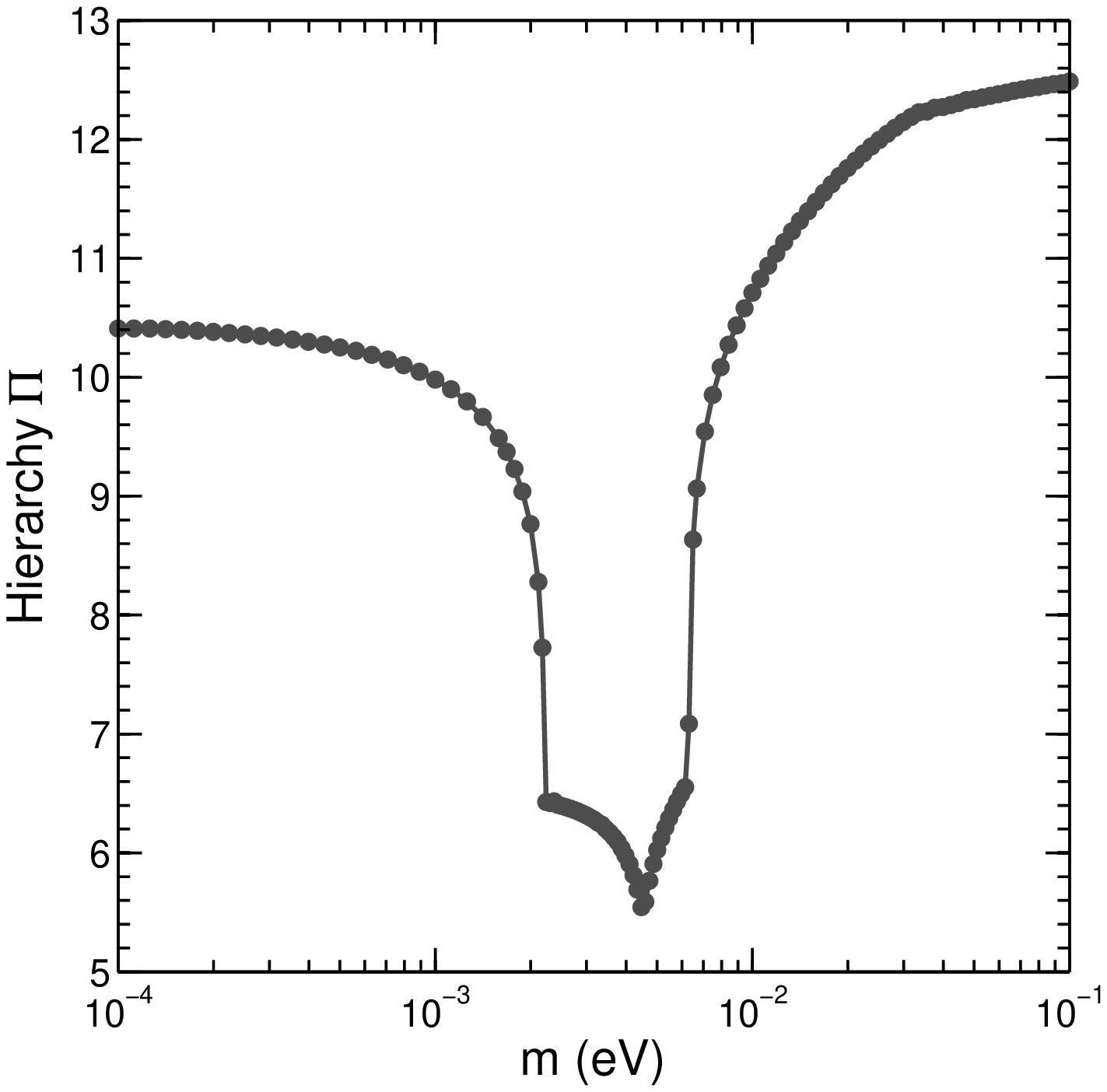}
\includegraphics[width=.45 \columnwidth]{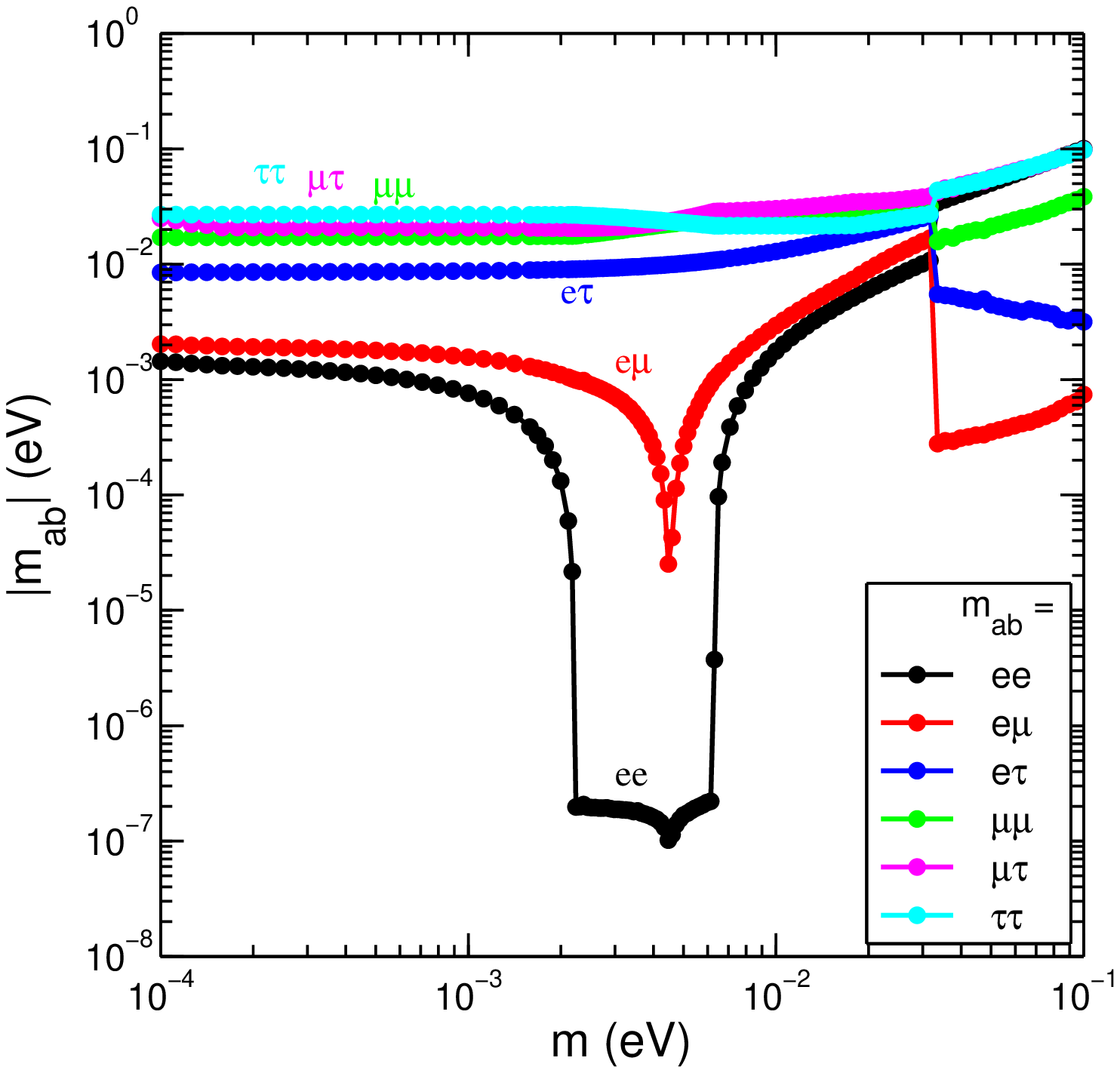}
}
\caption{\small LEFT: The minimal amount of Yukawa hierarchy $\Pi$ [defined in Eq. (\ref{tuning})] as a function of the lightest neutrino mass $m$. A clear minimum in the hierarchy appears in a narrow range of $m$, where $\Pi \simeq 5.5$ and $\left|m^{\nu}_{ee}\right| \simeq 10^{-7}$ eV. 
RIGHT: Values of the various neutrino mass entries $\left|m^{\nu}_{ab}\right|$ as a function of $m$. The parameters $\theta_{12}$, $\theta_{13}$, $\theta_{23}$, $\Delta m^2_{13}$, and $\Delta m^2_{21}$ are fixed to their best-fit measured values \cite{GonzalezGarcia:2012sz}.}
\label{Pimee}
\end{figure}

The degree of hierarchy in a specific ``Yukawa'' matrix $C^{(9)}_{ab}$ can be quantified by means of Eq.~(\ref{loopnumass}) and using the neutrino oscillation data as input for $m^{\nu}_{ab}$. 
We choose to define the hierarchy of $C^{(9)}_{ab}$ in the charged-lepton mass eigenstate basis. We look at the hierarchy among the absolute values of the neutrino mass matrix entrances\footnote[10]{Alternatively, we could have defined the hierarchy by means of the eigenvalues of $C^{(9)}_{ab}$, which yields different but qualitatively similar results. However, in that case the connection between $C^{(9)}_{ab}$, neutrino oscillation parameters, and charged-lepton masses is not apparent.} (note, however, that they are in general complex entries with their relative phase adjusted to agree with experimental data). The degree of hierarchy $\Pi$ is then defined as the sum of squared logarithms of the relative sizes of the Yukawa couplings:
\be
\label{tuning}
\Pi^2 = \underset{N}{\mathrm{min}} \sum_{a\geq b} \left[ \mathrm{log}
\left(N \frac{\left|m^{\nu}_{ab}\right|}{m_{l_a} m_{l_b}} 
\right) \right]^2\,.
\ee
The constant $N$ is minimized over to account for the fact that the definition of hierarchy should not be dependent on the overall $C^{(9)}_{ab}$ normalization. $\Pi$ in Eq.~(\ref{tuning}) then provides us with a way to quantify the amount of hierarchy required in $C^{(9)}_{ab}$ to be compatible with data from neutrino oscillation experiments. In Fig.~\ref{Pimee} (LEFT) we show the minimal amount of hierarchy $\Pi$ as a function of $m$, with the corresponding values of $\left|m^{\nu}_{ee}\right|$, $\left|m^{\nu}_{e\mu}\right|$ as well as the rest of $\left|m^{\nu}_{ab}\right|$ shown in Fig.~\ref{Pimee} (RIGHT). The neutrino oscillation parameters $\theta_{12}$, $\theta_{23}$, $\theta_{13}$, $\Delta m^2_{13}$, and $\Delta m^2_{21}$ are set to their best-fit values,
%%%%%%%%%%%%%%%% 
%%%%%%%%%%%%%%%%
and we use a global minimization\footnote[11]{Global minimization in multidimensional spaces is generically difficult.  We found the \texttt{MATLAB} tool box with \texttt{GlobalSearch} setup and the \texttt{fmincon} option for finding minima of constrained nonlinear multivariable functions suited our task well \cite{MATLAB}.} in the neutrino parameters $\delta$, $\alpha_1$, and $\alpha_2$, as well as in $N$. As shown in Fig.~\ref{Pimee}, the region with the  lowest hierarchy corresponds to $0.002 \, \mathrm{
eV} < m < 0.007 \, \mathrm{eV}$ (the ``chimney"), where $\left|m^{\nu}_{ee}\right| \sim 10^{-7}$ eV. The Yukawa hierarchy in this region is of the same order as the one already present in the SM 
charged-lepton mass pattern 
%%%%%%%%%%%%%%%% 
%\begin{figure}[t]
%\center{
%\includegraphics[width=.5 \columnwidth]{figs/hierarchy_3.eps}
%\includegraphics[width=.48 \columnwidth]{figs/hierarchy_1b.eps}
%}
%\caption{\small LEFT: Value of the different entries in the neutrino mass matrix $\left|m^{\nu}_{ab}\right|$ for the minimal hierarchy $\Pi$ as a function of the lightest neutrino mass $m$ (in the NO scenario). RIGHT: Value of the minimal hierarchy $\Pi$ (defined in Eq.~\ref{tuning}) as a function of $m$.} 
%\label{masses_matrix}
%\end{figure}
%%%%%%%%%%%%%%%%
%
\be
\label{tuning2}
\underset{N}{\mathrm{min}} \sqrt{\left[ \mathrm{log}
\left(N m_{l_e} 
\right) \right]^2+\left[ \mathrm{log}
\left(N m_{l_{\mu}} 
\right) \right]^2+\left[ \mathrm{log}
\left(N m_{l_{\tau}} 
\right) \right]^2} \sim 6\,.
\ee
We find that the hierarchy is globally minimized for $m \sim 0.0045$ eV, when both $\left|m^{\nu}_{ee}\right|$ and $\left|m^{\nu}_{e\mu}\right|$ reach their minimal value ($\left|m^{\nu}_{ee}\right| \sim 10^{-7}$ eV, $\left|m^{\nu}_{e\mu}\right| \sim 10^{-5}$ eV). The fact that from Fig.~\ref{Pimee} there is only one value of $m$ for which both $\left|m^{\nu}_{ee}\right| \rightarrow 0$ and $\left|m^{\nu}_{e\mu}\right| \rightarrow 0$ can be understood by noticing that if $m^{\nu}_{ee} = 0$ and $m^{\nu}_{e\mu} = 0$ then fixing the oscillation parameters $\theta_{12}$, $\theta_{13}$, $\theta_{23}$, $\Delta m^2_{13}$, and $\Delta m^2_{21}$ results in a unique prediction for $\delta$, $\alpha_1$, $\alpha_2$ and $m$ (recall the discussion from Sec.~\ref{sec:textures}).

%\medskip 
From the previous discussion, the texture with $m^{\nu}_{ee} \simeq 0$ and $m^{\nu}_{e\mu} \simeq 0$ emerges as the one with the least hierarchy in the $C^{(9)}_{ab}$ when being compatible with neutrino oscillation experiments. Nevertheless, as anticipated, an amount of hierarchy similar to the one present in the charged-lepton sector is required to satisfy the oscillation data.

%%%%%%%%%%%%%%%%%%%%%%%%%%%%%%%%%%%%%%%%%%%%%%%%%%%%%%%%%%%%
\subsection{Tree-level and loop-induced BSM completions of $\mathcal{O}^9$}
\label{sec:BSM}
%%%%%%%%%%%%%%%%%%%%%%%%%%%%%%%%%%%%%%%%%%%%%%%%%%%%%%%%%%%%

We now explore possible renormalizable Lagrangians that would generate the $\mathcal{O}^9$ operator in Eq.~(\ref{D9}).\footnote[12]{A classification of all possible renormalizable completions was recently given in Ref.~\cite{Aparici:2013xga}.}  The leptons in Eq.\ (\ref{D9}) form a bilinear $\overline{\ell^c}_{R_a}\, \ell_{R_b}$ that is Lorentz and $SU(2)_L$ invariant but has hypercharge $Y=-2$.  The simplest possible completion\footnote[13]{As shown in Ref.~\cite{Aparici:2013xga}, it is also possible to build renormalizable completions of $\mathcal{O}^9$ that do not involve the bilinear $\overline{\ell^c}_{R_a}\, \ell_{R_b}$ appearing explicitly in the Lagrangian. These completions, however, require a substantially larger amount of degrees of freedom, and we will not consider them in this work.} of $\overline{\ell^c}_{R_a}\, \ell_{R_b}$ then requires a scalar field $\rho^{++}$, $SU(2)_L$ singlet with hypercharge $Y=2$, to form the renormalizable and $SU(2)_L \times U(1)_Y$ gauge invariant term $C_{ab}\,\
overline{
\ell^c}_{R_a}\, \ell_{R_b}\,\rho^{++}$, with $C_{ab}$  a Yukawa coupling matrix in flavor space. As the field $\rho^{++}$ is a singlet under $SU(2)_L$, it does not couple directly to the $W$ bosons. The extra field(s) needed to connect $\rho^{++}$ to the $W$ bosons [{\it cf.}\@ Eq.\;(\ref{D9WW})] will in turn determine whether the nonrenormalizable operator $\mathcal{O}^9$ is induced at tree level or at loop level. 

\smallskip
For a tree-level completion, $\rho^{++}$ has to mix with another state which has nontrivial $SU(2)_L$ quantum numbers. The simplest possibility is to introduce a scalar $SU(2)_L$ triplet field $\Delta$ with 
hypercharge $Y=1$ \cite{Chen:2006vn,delAguila:2011gr}
\be
\label{triplet}
\Delta = \left(\begin{array}{cc}
\Delta^+/\sqrt{2} & \Delta^{++} \\
\Delta^0  & -\Delta^+/\sqrt{2}
\end{array}
\right)
\ee    
The $\Delta^{++}$ component of $\Delta$ may mix with $\rho^{++}$ upon electroweak symmetry breaking, giving rise to two doubly charged mass eigenstates $\Delta_{1,2}^{++}$ with masses $m_{\Delta_{1,2}}$. At the same time $\Delta^0$ has to develop a vacuum expectation value (\textit{vev}) $\left\langle \Delta^0 \right\rangle = v_{\Delta}$ to generate the coupling of $\Delta_{1,2}^{++}$ to $W$ bosons. Two possible ways to obtain the $\Delta^{++}$-$\rho^{++}$ mixing (parametrized via a mixing angle $\theta_{\Delta}$) have been explored. 

The first way is through an operator $H^{\dagger}\, \Delta\, \tilde{H} \,\rho^{++}$  (with $H$ being the SM scalar doublet and $\tilde{H} = i \sigma_2 H^*$) \cite{Chen:2006vn,Chen:2007dc}. In this case $\theta_{\Delta}$ may be sizable [$\mathrm{\sin(2\theta_{\Delta})} \sim v^2/m^2_{\Delta_{1,2}}$] if the masses $m_{\Delta_{1,2}}$ are close to the electroweak scale $v$. However, in this scenario an additional contribution to neutrino masses, coming from a type-II seesaw mechanism, cannot be consistently avoided (see the discussion in Ref.~\cite{delAguila:2011gr}). The appearance of this type-II seesaw contribution makes the mass generation from the effective operator $\mathcal{O}^9$ somewhat redundant.
The second way, proposed in Ref.~\cite{delAguila:2011gr}, overcomes this issue by  inducing the $\Delta^{++}$-$\rho^{++}$ mixing by an operator $\mathrm{Tr}\left[ \Delta^{\dagger}\Delta^{\dagger}\right]\,\rho^{++}$. A $Z_2$ symmetry, which needs to be spontaneously broken\footnote[14]{As discussed in Ref.~\cite{delAguila:2011gr}, the spontaneous breaking of this $Z_2$ symmetry leads to a domain-wall problem in this scenario.} in order for $\mathcal{O}^9$ to be generated at tree level, then forbids the terms in the Lagrangian that would generate a type-II seesaw contribution to neutrino masses. However, the mixing angle $\theta_{\Delta}$ is suppressed in this case [$\mathrm{\sin(2\theta_{\Delta})} \sim v_{\Delta}/m_{\Delta_{1,2}} \ll 1$] due to the smallness of  the triplet \textit{vev} $v_{\Delta}$, as needed to satisfy electroweak precision constraints ($v_{\Delta} \lesssim 5$\;GeV $\ll v$).

\smallskip
An elegant solution to the above problems for tree-level completions is obtained by considering renormalizable completions at loop level. No mixing of the $\rho^{++}$ field is necessary, but instead the $\rho^{++}$ coupling to $W$ bosons appears at 1 loop (or higher) via new fields charged under both $SU(2)_L$ and $U(1)_Y$. If the new fields respect a $Z_2$ symmetry, which remains unbroken after electroweak symmetry  breaking, this guarantees that neutrino masses will first appear at 3 (or more) loops.  Also, because of the $Z_2$ symmetry, these scenarios could automatically incorporate a stable dark matter candidate. A concrete realization of this idea is given in Ref.~\cite{Gustafsson:2012vj}, which together with the discussed $\rho^{++}$ scalar field introduces two new  fields which are odd under the $Z_2$ symmetry: an inert scalar $SU(2)_L$ doublet,
\be
\label{doublet}
H_2 = \left(\begin{array}{c}
\Lambda^{+} \\
H_{0}+ i\,A_0
\end{array}
\right), %\quad \quad \quad S^{+}
\ee   
and a scalar $SU(2)_L$ singlet, $S^+$, with hypercharge $Y=1$. Altogether, the LNV part of the Lagrangian is given by 
\bea
\label{Lcocktail}
-\mathcal{L}_{LNV} =& C_{ab} \, 
\overline{\ell^{c}}_{R_a} \ell_{R_b} \, \rho^{++} + \frac{\lambda_5}{2} \left(H^{\dagger}_{1} H_2 \right)^2 
+ \kappa_1 \, H_2^{T} i \sigma_2 H_{1}\, S^{-}  \nonumber \\ 
\, & + \,\kappa_2 \, \rho^{++} S^{-} S^{-} + \,\xi \, H_2^{T} i \sigma_2 H_{1}\, S^{+} \, \rho^{--} + \mathrm{h.c.} \quad \quad 
\eea
Because of the $\kappa_1$-coupling in Eq.~(\ref{Lcocktail}), the states $S^{+}$ and $\Lambda^{+}$ mix upon electroweak symmetry breaking (the mixing angle being $\theta^{+}$), giving rise to two charged mass eigenstates $H^+_{1,2}$ with masses $m_{H^+_{1,2}}$. The states $H^+_{1,2}$, $A_0$, and $H_0$  are thus the only particles with negative $Z_2$ parity ($\rho^{++}$ has positive $Z_2$ parity), so if the lightest of these states is either $A_0$ or $H_0$ the setup will automatically provide a viable dark matter candidate. In this model, the unbroken $Z_2$ symmetry results in the $\mathcal{O}^{9}$ operator being generated at 1-loop level, with neutrino masses appearing at 3 loops (see Ref.~\cite{Gustafsson:2012vj} for details). We want to stress that, as for the tree-level case, various (possibly many) loop realizations of the $\mathcal{O}^{9}$ operator may be possible.

\medskip

A common feature of all the scenarios discussed above is the presence of the $\rho^{++}$ scalar particle, coupling to right-handed charged leptons through a term $C_{ab}\,\overline{\ell^c}_{R_a}\, \ell_{R_b}\,\rho^{++}$. This scalar particle then mediates lepton flavor violation (LFV) processes such as $\mu^{+}\rightarrow e^{+} e^{+} e^{-}$, $\tau^{-}\rightarrow e^{+} e^{-} e^{-}$, or  $\tau^{-}\rightarrow e^{+} \mu^{-} \mu^{-}$ at tree level (with their amplitudes $a \rightarrow b \,c\, d$ proportional to $\left| C_{ab}^{*}C_{cd}\right|/m_{\rho}^2$), and $\mu^{+}\rightarrow e^{+}\, \gamma$ at 1 loop. The experimental constraints on these processes are rather stringent, especially for the very rare decays $\mu^{+}\rightarrow e^{+} e^{+} e^{-}$ \cite{Bellgardt:1987du},  $\mu^{+}\rightarrow e^{+}\, \gamma$ \cite{Adam:2013mnn}, and $\tau^{-}\rightarrow e^{+} \mu^{-} \mu^{-}$ \cite{Hayasaka:2010np}. The scenarios discussed above can have rates close to the current experimental limits. While a $\mu^{+}\rightarrow 
e^{+} e^{+} e^{-}$ signal could be avoided by tuning  $C_{ee} C_{e\mu}$ to be tiny, the LFV processes $\mu^{+}\rightarrow e^{+}\, \gamma$ and $\tau^{-}\rightarrow e^{+} \mu^{-} \mu^{-}$ are predicted close to the current experimental sensitivity since the Yukawa couplings $C_{e\tau}$, $C_{\mu\mu}$, and $C_{\mu\tau}$ cannot be very small in these scenarios in order to reproduce the correct values for the observed neutrino mass spectrum \cite{delAguila:2011gr,Gustafsson:2012vj}.  A detailed analysis of LFV constraints for these scenarios lies beyond the scope of this paper (see Ref.~\cite{Nebot:2007bc} for a study of LFV constraints on processes mediated by a $\rho^{++}$ singlet scalar). Relevant for the next section are, however, the limits on $C_{ee}$ for the specific benchmark scenarios discussed in Ref.~\cite{delAguila:2011gr} ($C_{ee} \lesssim \sqrt{4\pi}$) and Ref.~\cite{Gustafsson:2012vj} ($C_{ee} \lesssim 0.1$).

\medskip

The discussion of this section highlights the feasibility of finding renormalizable completions of $\mathcal{O}^{9}$, both at tree level and 1-loop level, and the main features of these completions. As discussed previously, the smallness of neutrino masses compared to the electroweak scale is  explained in these scenarios in terms of a 2- or 3-loop suppression. Simultaneously, approximate texture zeroes appearing in the neutrino mass matrix give rise to specific correlations among the various neutrino oscillation parameters, which are perfectly compatible with current experimental data and will be tested in the near future. Furthermore, the $0\nu \beta\beta$ decay process in these models has particular features, which we study in detail in the next section.

%%%%%%%%%%%%%%%%%%%%%%%%%%%%%%%%%%%%%%%%%%%%%%%%%%%%%%%%%%%%
\section{Neutrinoless double $\beta$ decay ($0\nu\beta\beta$)} \label{sec:0nu2beta}
%%%%%%%%%%%%%%%%%%%%%%%%%%%%%%%%%%%%%%%%%%%%%%%%%%%%%%%%%%%%

\begin{figure}[t]
\center{\includegraphics[width=.35 \columnwidth]{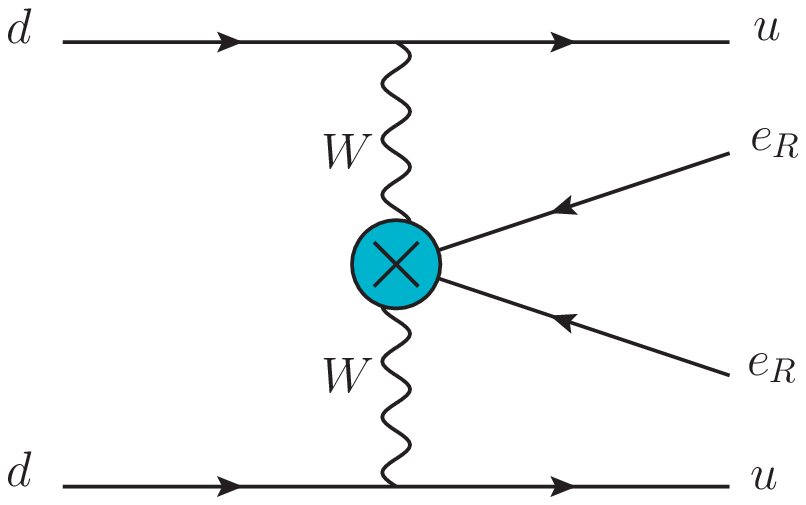}}
\vspace{4mm}
\center{\includegraphics[width=.35 \columnwidth]{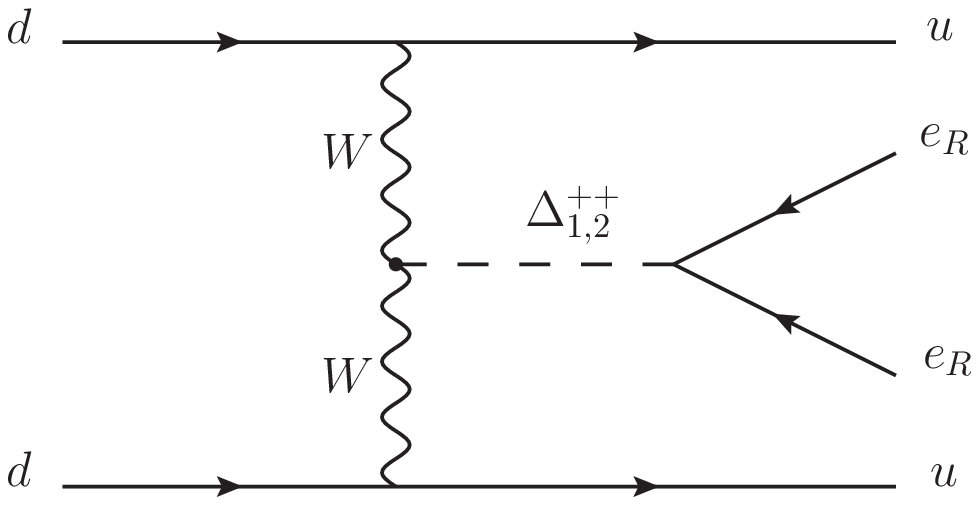}
\hspace{4mm}
\includegraphics[width=.33 \columnwidth]{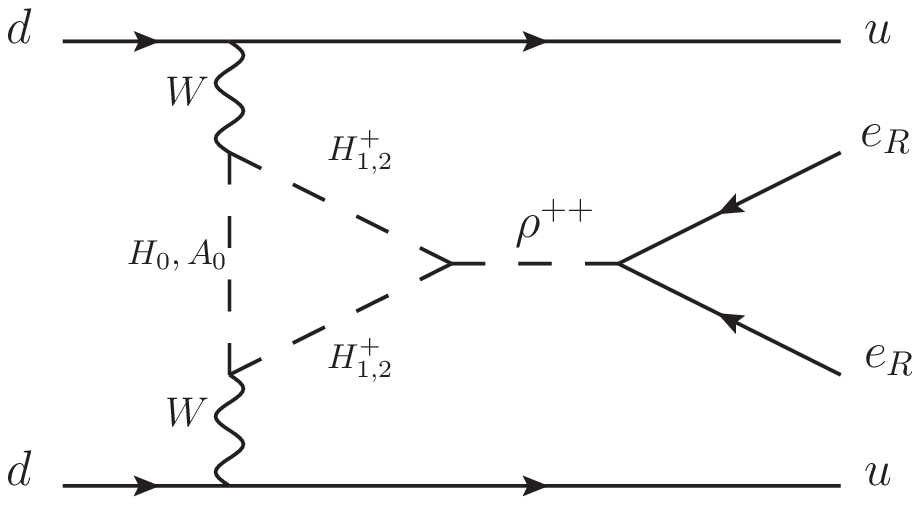}}
\caption{\small 
TOP: Effective $0\nu\beta\beta$ contribution from $\mathcal{O}^9$. 
BOTTOM LEFT: $0\nu\beta\beta$ from tree-level BSM completions of $\mathcal{O}^9$. 
BOTTOM RIGHT: $0\nu\beta\beta$ from 1-loop BSM completions of $\mathcal{O}^9$.} \label{fig:0nu2beta}
\end{figure}

Scenarios where LNV new physics couples directly only to right-handed charged leptons are very interesting for the potential $0\nu\beta\beta$ decay of atomic nuclei. As discussed in the previous section, the operator $\mathcal{O}^9$ in Eq.~(\ref{D9})  leads to a neutrino mass matrix $m^{\nu}_{ab}$ of the form (\ref{loopnumass}) which then generically contains approximate zeroes $m^{\nu}_{ee} \simeq 0$, $m^{\nu}_{e\mu} \simeq 0$. 
As a result, the contribution from light neutrinos to the $0\nu\beta\beta$ decay amplitude, being proportional to $\mathcal{A}^{\nu}_{0\nu\beta\beta} \sim m^{\nu}_{ee}/\left\langle p^2 \right\rangle$ with $\left\langle p^2 \right\rangle \sim (100 \, \mathrm{MeV})^2$ the square of the typical momentum transfer between nucleons in the decay process \cite{Blennow:2010th}, is extremely suppressed due to the smallness of $m^{\nu}_{ee}$. Meanwhile, the contribution from the short-distance physics encoded in the operator $\mathcal{O}^9$ contributes to the $0\nu\beta\beta$ decay (as shown in Fig.~\ref{fig:0nu2beta}) without suffering from the 2-loop and $m^2_{l_e}$ suppression that affects the light-neutrino exchange via the $m^{\nu}_{ee}$ factor [recall Fig.~\ref{fig:TwoLoops} and Eq.~(\ref{loopnumass})].
%%%%%%%%%%%%%%%% 
%%%%%%%%%%%%%%%%

In these scenarios, the short-distance contribution is thus the dominant source of $0\nu\beta\beta$ decays. This is in contrast with more conventional scenarios of neutrino mass generation, such as seesaw models \cite{LopezPavon:2012zg}. We note that if LNV physics couples directly to quarks, or if the SM gauge group is extended, it is also generically possible for the short-range contributions to $0\nu\beta\beta$ decay to dominate (see Ref.~\cite{Bonnet:2012kh} for a discussion). Depending on the value of $C_{ee}$ (the Yukawa coupling from the renormalizable BSM theory, as discussed in Sec.~\ref{sec:BSM}), this contribution could potentially be large enough to make the process detectable in ongoing and future $0\nu\beta\beta$ decay experiments, including  GERDA\cite{Schonert:2005zn,Agostini:2013mzu}, EXO\cite{Akimov:2005mq}, SNO+\cite{Chen:2005yi}, KamLAND-Zen\cite{Mitsui:2011zza,KamLANDZen:2012aa}, CUORE\cite{Arnaboldi:2002du}, NEXT\cite{Dafni:2012nwa,GomezCadenas:2012jv}, 
MAJORANA \cite{Gaitskell:2003zr} and SuperNEMO\cite{Arnold:2010tu}.

\vspace{3mm}

The effective six-fermion contact interaction relevant for $0\nu\beta\beta$ decays, obtained from the short-distance contribution encoded in $\mathcal{O}^9$, can be written as (see Refs.~\cite{Pas:2000vn,Bergstrom:2011dt}) 
\be
\label{Lbetabeta}
\mathcal{L}_{0\nu\beta\beta} =  \frac{G_F^2}{2\, m_p}\, \epsilon_3 \, J^{\mu}\,J_{\mu}\, \bar{e}(1-\gamma_5)e^c.
\ee  
with the vector-axial hadronic currents being $J^{\mu} = \bar{u}\gamma^{\mu}(1-\gamma_5)d$. The parameter $\epsilon_3$ encodes the information on the underlying BSM LNV physics, and may be written 
in terms of the Feynman amplitude $\mathcal{A}^{\mathrm{SD}}_{0\nu\beta\beta}$ for the short-distance $0\nu\beta\beta$ process 
\be
\label{E3_1}
\epsilon_3 = - 2\,m_p\, \mathcal{A}^{\mathrm{SD}}_{0\nu\beta\beta}\, .
\ee 

The relevant measurable quantity is the half-life $T^{0\nu\beta\beta}_{1/2}$ due to $0\nu\beta\beta$ decays for various nuclei that can undergo a $2\beta$ decay. The nonobservation of $0\nu\beta\beta$ decays sets experimental lower bounds on the values of $T^{0\nu\beta\beta}_{1/2}$ for the different nuclei. 
In principle, contributions from all possible sources of $0\nu\beta\beta$ decay have to be considered coherently. However, in scenarios where one source is dominant, the analysis simplifies considerably. Assuming, respectively, that the dominant contribution to the $0\nu\beta\beta$ process comes from the light-neutrino exchange or from the short-distance physics leading to Eq.~(\ref{Lbetabeta}), the half-life $T^{0\nu\beta\beta}_{1/2}$ is very well approximated by either 
\be
\label{HalfLife}
\left[ T^{0\nu\beta\beta}_{1/2} \right]^{-1} \simeq G_{01}\, \frac{|m^{\nu}_{ee}|^2}{m^2_{l_e}}\, |\mathcal{M}^{\mathrm{\nu}}|^2
\quad \quad \mathrm{or} \quad \quad 
\left[ T^{0\nu\beta\beta}_{1/2} \right]^{-1} \simeq G_{01}\, |\epsilon_3|^2\, |\mathcal{M}^{\mathrm{SD}}|^2. 
\ee 
Here $G_{01}$ is a phase-space factor, characteristic of the specific decaying nucleus considered \cite{Suhonen:1998ck}, and $\mathcal{M}^{\mathrm{SD}}$ (short-distance)  and $\mathcal{M}^{\mathrm{\nu}}$ (light-neutrino exchange) are the nuclear matrix elements (NMEs) for each specific nucleus. Equation~(\ref{HalfLife}) then allows us to derive limits on $m^{\nu}_{ee}$ or $\epsilon_3$ from experimental bounds on $T^{0\nu\beta\beta}_{1/2}$. The values of  $G_{01}$, $\mathcal{M}^{\mathrm{\nu}}$ and $\mathcal{M}^{\mathrm{SD}}$ are taken from \cite{Suhonen:1998ck}, \cite{GomezCadenas:2010gs} and \cite{Deppisch:2012nb}, respectively,  and are presented in Table~\ref{tab:1}. 
%%%%%%%%%%%%%%%%%%%%%%%%%%%%%%%%%%%%%%%%%%
\begin{table*}[]
\begin{center}
\begin{tabular}{rcc c  }
\hline\hline \\[-4mm]
& $G_{01}\,\, (10^{-14}\, \mathrm{yr}^{-1})$ & $\left|\mathcal{M}^{\nu} \right|$ &  $\left|\mathcal{M}^{\mathrm{SD}} \right|$  \\[0.5mm]
\hline
$\,^{76}\mathrm{Ge}$ & 0.623 & 4.07 & 213 \cr
$\,^{136}\mathrm{Xe}$ & 4.31 & 2.82 & 109 \cr
$\,^{150}\mathrm{Nd}$ & 19.2 & 2.33 & 311 \cr
$\,^{130}\mathrm{Te}$ & 4.09 & 3.63 & 198 \cr
$\,^{82}\mathrm{Se}$ & 2.70 & 3.48 & 192 \cr
\hline\hline
\end{tabular}
\end{center}
\caption{\small Values of the phase-space factors $G_{01}$ for the different nuclei, taken from Ref.~\cite{Suhonen:1998ck}.  The nuclear matrix elements  (NMEs) $\left|\mathcal{M}^{\nu} \right|$ are the averaged best estimates from Ref.~\cite{GomezCadenas:2010gs}. The NMEs $\left|\mathcal{M}^{\mathrm{SD}} \right|$ are the values given in Ref.~\cite{Deppisch:2012nb} from their computation using the proton-neutron quasiparticle random phase approximation (pn-QRPA) approach.}
\label{tab:1}
\end{table*}
%%%%%%%%%%%%%%%%%%%%%%%%%%%%%%%%%%%%%%%%%%
Note, however, that different methods of evaluating NMEs can give results differing by a factor $\sim\, 2$, emerging from the different  approximations used by each method and the propagation of their respective uncertainties. This induces a corresponding uncertainty in the derived limits on $\epsilon_3$ and $m^{\nu}_{ee}$ quoted below (see Refs.~\cite{Blennow:2010th,Pas:2000vn,Bergstrom:2011dt,GomezCadenas:2010gs} for more details including a discussion of the theoretical uncertainties).

\medskip
As discussed above, in scenarios where $\mathcal{O}^9$ constitutes the leading source of LNV, the short-distance contribution to $0\nu\beta\beta$ decay largely dominates over the light-neutrino exchange one. 
As an illustration, we will now consider the tree-level \cite{Chen:2006vn,delAguila:2011gr} and 1-loop \cite{Gustafsson:2012vj} renormalizable completions to $\mathcal{O}^9$ discussed in Sec.~\ref{sec:BSM} (see Fig.~\ref{fig:0nu2beta}). For the tree-level case, the $0\nu\beta\beta$ decay amplitude $\mathcal{A}^{\mathrm{SD}}_{0\nu\beta\beta}$ reads
\be
\label{E3_2}
\mathcal{A}^{\mathrm{tree}}_{0\nu\beta\beta} = \mathrm{s}_{2\theta_{\Delta}}\,v_{\Delta} \, 
\frac{\left(m^2_{\Delta_1} - m^2_{\Delta_2} \right)}{m^2_{\Delta_1}\, m^2_{\Delta_2}}  \, C_{ee}.
\ee  
As expected, $\mathcal{A}^{\mathrm{tree}}_{0\nu\beta\beta}$ vanishes in the limit $m^2_{\Delta_1} = m^2_{\Delta_2}$, since in this limit the $\Delta^{++}-\rho^{++}$ mixing term is absent from the Lagrangian and the lepton number would be exactly conserved  (recall the discussion in Sec.~\ref{sec:BSM}). For the 1-loop case, the $0\nu\beta\beta$ decay amplitude $\mathcal{A}^{\mathrm{SD}}_{0\nu\beta\beta}$ can be computed in a straightforward way to give  
\be
\label{E3_3}
\begin{array}{cl}
\mathcal{A}^{\mathrm{loop}}_{0\nu\beta\beta} = & \frac{\Delta m^2_+\,\, \mathrm{s}_{2\theta^+}}{8 \pi^2\,\,m^2_{\rho}} \,C_{ee} \,\, \times \\
&\left\lbrace \left[\kappa_2 \, \Delta m^2_+\, \mathrm{s}_{2\theta^+} - 
\xi\,v\, (\mathrm{c}^2_{\theta^+}\,m^2_{H^+_2}+\mathrm{s}^2_{\theta^+}\,m^2_{H^+_1})\right]
\left[F_{H^+_1,H^+_2,H_0} - F_{H^+_1,H^+_2,A_0}\right] \right.\\
& \left. - \xi\,v \left[ m^2_{H_0}\,F_{H^+_1,H^+_2,H_0} - m^2_{A_0}\,F_{H^+_1,H^+_2,A_0}\right] \right\rbrace
\end{array}
\ee
with $\Delta m^2_+ = m^2_{H^+_1} - m^2_{H^+_2}$ and
\be
F_{a,b,c} = \int_0^1 d\,x \int_0^{1-x} d\,y \, \frac{x\,y}{\left(x\,m^2_a + y\,m^2_b + (1-x-y)\,m^2_c\right)^2}\,.
\ee
It can then be verified, using Eqs.~(\ref{E3_2}) and (\ref{E3_3}), that both for the tree-level and 1-loop renormalizable completions $\mathcal{A}^{\mathrm{SD}}_{0\nu\beta\beta} \gg \mathcal{A}^{\nu}_{0\nu\beta\beta}$.  

\medskip
Using the benchmark scenario from Ref.~\cite{delAguila:2011gr} ($m_{\Delta_1} = 10$ TeV, $m_{\Delta_2} = 2$ TeV, $v_{\Delta} = 2$ GeV and $\mathrm{s}_{2\theta_{\Delta}}$ = 0.00125) and the benchmark scenario from Ref.~\cite{Gustafsson:2012vj} ($m_{H_0} = 70$ GeV, $m_{A_0} = 475$ GeV, $m_{H^+_1} = 90$ GeV, $m_{H^+_2} = 850$ GeV, $m_{\rho} = 2$ TeV, $\kappa_{2} = 3$ TeV, $\xi = -2.5$, and $\mathrm{s}_{2\theta^{+}} = 1$), we, respectively, obtain for $\epsilon_3$
\be
\label{E3bis}
\epsilon^{\mathrm{tree}}_{3}  
= 1.1 \times 10^{-9}\, \left| C_{ee} \right| \, \, \quad , \quad \quad 
\epsilon^{\mathrm{loop}}_{3}  = 1.2 \times 10^{-5}\, \left| C_{ee} \right| \, .
\ee
The predicted values of $\epsilon_3$ may vary depending on the specific tree-level/loop completion of $\mathcal{O}^9$ and the concrete values of the parameters within that scenario. However, Eq.~(\ref{E3bis}) may be regarded as a rough guide of what is to be generically expected. We note at this point that for $C_{ee} \rightarrow 0$ the short-distance contribution to $0\nu\beta\beta$ vanishes along with 
$m^{\nu}_{ee}$. However, for sizable values of $C_{ee}$, the short-distance contribution can indeed be very large (see the discussion below).

\medskip
The fact that $\epsilon^{\mathrm{loop}}_{3} \gg \epsilon^{\mathrm{tree}}_{3}$ in Eq.~(\ref{E3bis}) can be understood in the following way: The tree-level amplitude roughly scales as $\mathcal{A}^{\mathrm{tree}}_{0\nu\beta\beta} \propto  \mathrm{s}^2_{2\theta_{\Delta}}/m$, while the 1-loop amplitude scales as $\mathcal{A}^{\mathrm{loop}}_{0\nu\beta\beta} \propto 1/(4\pi)^2 \times \mathrm{s}^2_{2\theta^{+}}/m'$ with the mass scales $m$ and $m'$ being both $\mathcal{O}$(TeV) for the benchmark scenarios considered in Refs.~\cite{delAguila:2011gr} and \cite{Gustafsson:2012vj}. However, the mixing angle in the tree-level case is proportional to 
$v_{\Delta}$ and thus very constrained by electroweak precision observables ($\mathrm{s}_{2\theta_{\Delta}} \sim 0.001$), while in the 1-loop case, the relevant mixing angle is not severely constrained and can be sizable ($\mathrm{s}_{2\theta^{+}} \sim 1$). This may then result in $\mathcal{A}^{\mathrm{loop}}_{0\nu\beta\beta} \gg \mathcal{A}^{\mathrm{tree}}_{0\nu\beta\beta}$ despite the additional loop suppression factor $1/(4\pi)^2$.

%%%%%%%%%%%%%%%%
\begin{figure}[nt!]
\center{
\includegraphics[width=.78 \columnwidth]{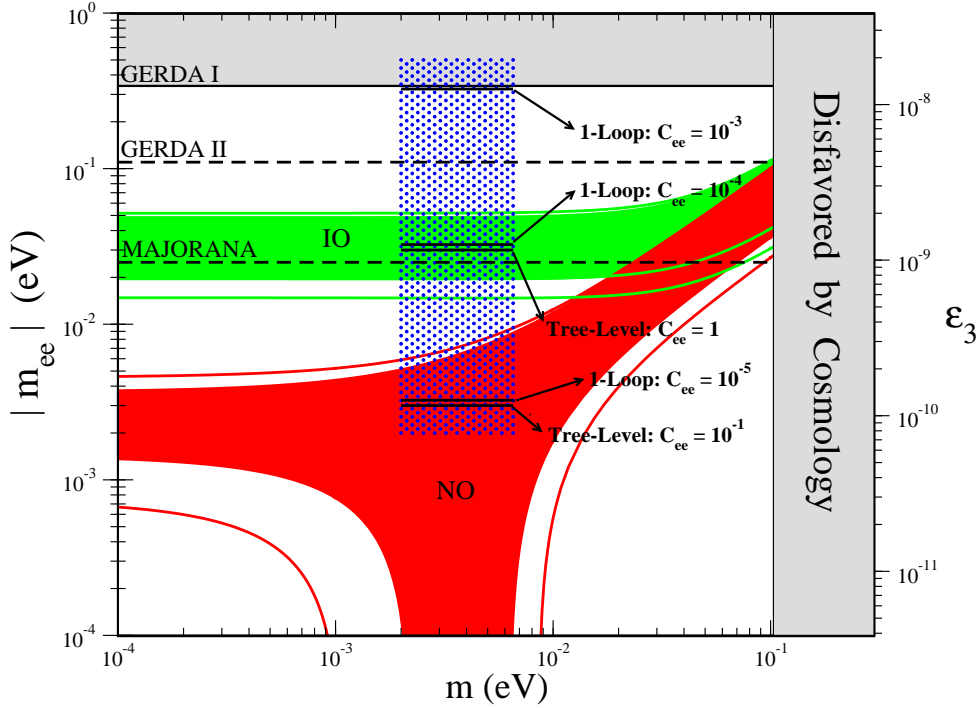}
}
\caption{\small Current limits and future sensitivities on $\epsilon_3$ and $\left| m^{\nu}_{ee} \right|$  from the $\,^{76}\mathrm{Ge}$ experiments GERDA and  MAJORANA (horizontal dashed-black lines), derived from $T^{0\nu\beta\beta}_{1/2}$ limits  with $0\nu\beta\beta$ decay dominated by either short-distance physics ($\epsilon_3$) or light-neutrino exchange ($\left| m^{\nu}_{ee} \right|$).  Allowed $\left| m^{\nu}_{ee} \right|$ regions for normal ordering NO (red) and inverted ordering IO (green) are shown for  neutrino oscillation parameters fixed to best-fit values (filled regions) and if allowed to vary within their $3\sigma$ uncertainty range (solid lines). The rectangle (blue dots) represents a generic region in the ($m,\epsilon_3$) plane for scenarios featuring the short-distance $\mathcal{O}^9$ operator in Eq.~(\ref{D9}) (recall that these scenarios have NO with $\left| m^{\nu}_{ee} \right| \simeq 0$). Its enclosed horizontal solid-black lines are concrete Yukawa choices  $C_{ee} =10^{-3},10^{-4},
 10^{-5}$ (1 loop) and $C_{ee} =1,10^{-1}$ (tree level) for the tree-level/1-loop renormalizable completion of $\mathcal{O}^9$ discussed in the text. The grey regions are currently excluded by cosmological observations and present $0\nu\beta\beta$ decay limits.} 
\label{FigGe}
\end{figure}
%%%%%%%%%%%%%%%%
%%%%%%%%%%%%%%%%
\begin{figure}[nt!]
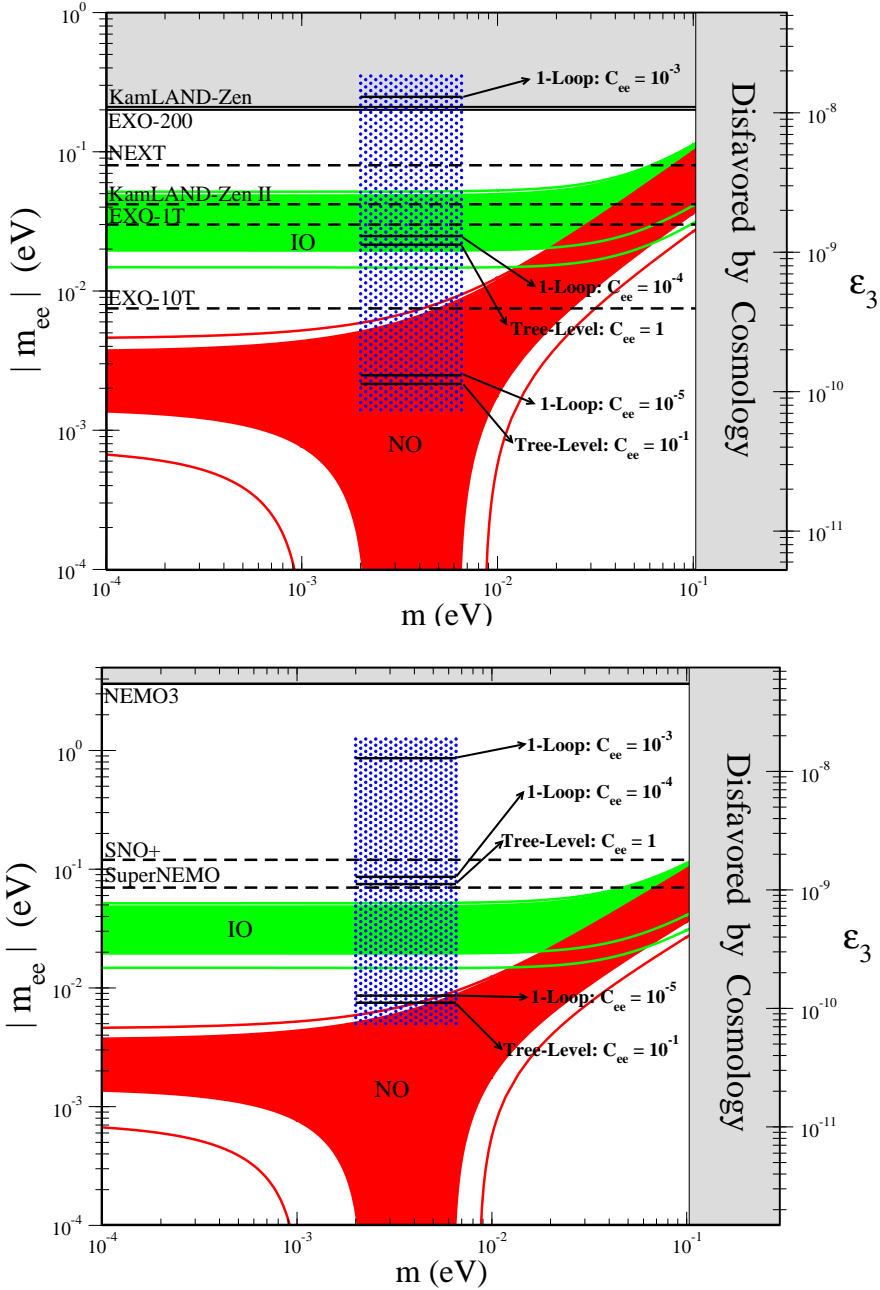

\center{
\includegraphics[width=.7 \columnwidth]{figs/M_eeNZeroXe.eps}

\vspace{4mm}

\includegraphics[width=.7 \columnwidth]{figs/M_eeNZeroNd.eps}
}
\caption{\small Same as Fig.~\ref{FigGe}, for the $\,^{136}\mathrm{Xe}$ experiments EXO, NEXT, and KamLAND-Zen (TOP) and the $\,^{150}\mathrm{Nd}$ experiments NEMO3, SNO+, and SuperNEMO (BOTTOM).} 
\end{figure}
%%%%%%%%%%%%%%%%
%%%%%%%%%%%%%%%%
\begin{figure}[nt!]
\center{
\includegraphics[width=.8 \columnwidth]{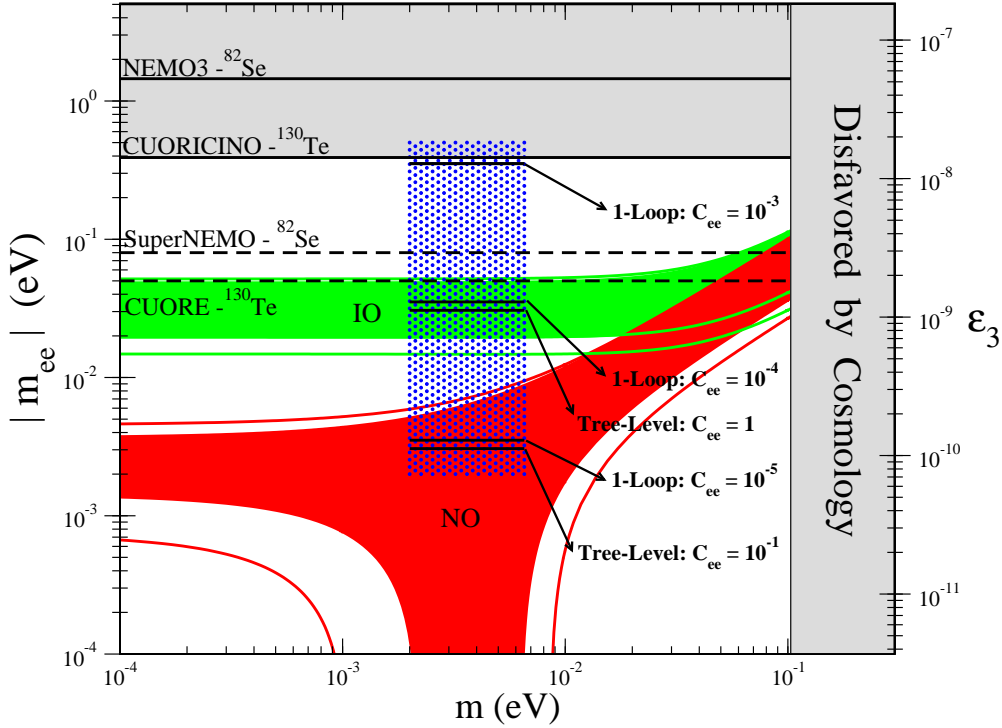}
}
\caption{\small Same as Fig.~\ref{FigGe}, for the $\,^{130}\mathrm{Te}$ experiments CUORICINO and CUORE and the $\,^{82}\mathrm{Se}$ experiments NEMO3 and SuperNEMO. Note that $\,^{130}\mathrm{Te}$ and $\,^{82}\mathrm{Se}$ can be presented in the same plot, as their $|\mathcal{M}^{\mathrm{SD}}|/|\mathcal{M}^{\nu}|$ ratios are numerically very similar (see the text and Table 1).
\vspace{4mm}}
\label{FigTeSe}
\end{figure}
%%%%%%%%%%%%%%%%

\bigskip
We can compare the predictions in Eq.~(\ref{E3bis}) with the current best experimental limits and future sensitivity prospects for five different nuclei that can undergo $2\beta^{-}$ decays: $\,^{76}\mathrm{Ge}$, $\,^{136}\mathrm{Xe}$, $\,^{150}\mathrm{Nd}$, $\,^{130}\mathrm{Te}$, and $\,^{82}\mathrm{Se}$. In the following we detail the present and future experimental bounds on $T^{0\nu\beta\beta}_{1/2}$ for these nuclei:
%%%%%%%%%%%%%%%%%%%%%%%%%%%%%%%%%%%%%%%%%%%%%%%%
\begin{itemize}
\item $\,^{76}\mathrm{Ge}$: The recent results from the phase~I run of GERDA \cite{Agostini:2013mzu,Macolino:2013ifa} place  the bound $T^{0\nu\beta\beta}_{1/2} > 2.1 \times 10^{25} \, \mathrm{yr}$ at $90 \,\%$ C.L., while phase II aims for a sensitivity of $T^{0\nu\beta\beta}_{1/2} > 2 \times 10^{26} \, \mathrm{yr}$ \cite{Schonert:2005zn}. MAJORANA plans to have an ultimate sensitivity of $T^{0\nu\beta\beta}_{1/2} > 4 \times 10^{27} \, \mathrm{yr}$ \cite{Gaitskell:2003zr} for the same nucleus after $\sim 10$ years of running.
 \item $\,^{136}\mathrm{Xe}$: The first results from the EXO-200 experiment set a  bound $T^{0\nu\beta\beta}_{1/2} > 1.6 \times 10^{25} \, \mathrm{yr}$ at $90 \,\%$ C.L.\cite{Auger:2012ar}. The  sensitivity prospects for the EXO-1T upgrade is $T^{0\nu\beta\beta}_{1/2} > 8 \times 10^{26} \, \mathrm{yr}$,  and $T^{0\nu\beta\beta}_{1/2} > 1.3 \times 10^{28} \, \mathrm{yr}$ for the ultimate EXO-10T upgrade \cite{Wamba:2005hr}. For the same nucleus, the current bound from the KamLAND-Zen experiment is $T^{0\nu\beta\beta}_{1/2} > 1.9 \times 10^{25} \, \mathrm{yr}$ at $90 \,\%$ C.L.\cite{KamLANDZen:2012aa,Gando:2012zm}, with a future planned sensitivity of $T^{0\nu\beta\beta}_{1/2} > 4 \times 10^{26} \, \mathrm{yr}$. The NEXT experiment aims to reach a sensitivity of $T^{0\nu\beta\beta}_{1/2} > 10^{26} \, \mathrm{yr}$ in the near future \cite{GomezCadenas:2012jv,GomezCadenas:2011it}. 
\item $\,^{150}\mathrm{Nd}$: The present bound from NEMO3 is $T^{0\nu\beta\beta}_{1/2} > 1.8 \times 10^{22} \, \mathrm{yr}$ at $90 \,\%$ C.L.\cite{Argyriades:2008pr}. The NEMO3 upgrade, SuperNEMO, aims to reach a sensitivity of $T^{0\nu\beta\beta}_{1/2} > 5 \times 10^{25}\, \mathrm{yr}$ \cite{Arnold:2010tu}.  At the same time, the SNO+ experiment expects, after 4 years of data taking, to reach a sensitivity of $T^{0\nu\beta\beta}_{1/2} > 1.6 \times 10^{25}\, \mathrm{yr}$ \cite{Maneira:2013fsa}.
\item $\,^{130}\mathrm{Te}$: The strongest bound for this nucleus is set by the CUORICINO experiment,  $T^{0\nu\beta\beta}_{1/2} > 3 \times 10^{24} \, \mathrm{yr}$ at $90 \,\%$ C.L.\cite{Arnaboldi:2008ds}. The CUORE experiment will substantially improve it, aiming for a sensitivity of $T^{0\nu\beta\beta}_{1/2} > 2 \times 10^{26}\, \mathrm{yr}$ in 5 years of data taking \cite{Arnaboldi:2002du}. 
\item $\,^{82}\mathrm{Se}$: The strongest bound for this nucleus is currently set by NEMO3 with $T^{0\nu\beta\beta}_{1/2} > 3.6 \times 10^{23} \, \mathrm{yr}$ at $90 \,\%$ C.L.\cite{Arnold:2005rz,Barabash:2010bd}. The planned sensitivity of SuperNEMO is $T^{0\nu\beta\beta}_{1/2} > 1.2 \times 10^{26}\, \mathrm{yr}$ \cite{Arnold:2010tu}. 
\end{itemize}
%%%%%%%%%%%%%%%%%%%%%%%%%%%%%%%%%%%%%%%%%%%%%%%%

A comparison between these current experimental limits and future expected sensitivities on $T^{0\nu\beta\beta}_{1/2}$ are presented as limits on $\epsilon_3$ and $m^{\nu}_{ee}$  in Figs.~\ref{FigGe}-\ref{FigTeSe}, for all the different nuclei discussed above. The predicted values of $\epsilon_3$ for the tree-level and 1-loop completions  of $\mathcal{O}^9$ (for different choices of $\left| C_{ee} \right|$), as well as the allowed region of $m^{\nu}_{ee}$ for NO and IO are also presented in the figures. It should be noted that the ratios $\left|m^{\nu}_{ee}\right|/\epsilon_3$ are different in these figures. This is due to the fact that for each nuclei the ratio $|\mathcal{M}^{\mathrm{SD}}|/|\mathcal{M}^{\nu}|$ is different (see Table 1).  For $\,^{130}\mathrm{Te}$ and $\,^{82}\mathrm{Se}$, we have numerically very similar  ratios of their NMEs $|\mathcal{M}^{\mathrm{SD}}|/|\mathcal{M}^{\nu}|$, and experiments involving these nuclei are therefore displayed together in Fig.~\ref{FigTeSe}.

\smallskip
For the scenarios discussed in this work, with LNV induced by the effective operator $\mathcal{O}^9$, the short-distance contribution to $0\nu\beta\beta$ decays is by far dominant over long-range neutrino exchange (for $\left| C_{ee} \right| = 10^{-3}$ the value $\left| m^{\nu}_{ee} \right|$ is $\sim10^{-8}$ eV for the 1-loop scenario in Ref.~\cite{Gustafsson:2012vj} and does not even appear in the range covered by the 
Figs.~\ref{FigGe}-\ref{FigTeSe}).
It is seen that for a not too small value of $\left| C_{ee} \right|$ the short-distance contribution to $0\nu\beta\beta$ decays could be within reach by both current and future experiments, in particular for 1-loop completions of $\mathcal{O}^9$ (for the specific 1-loop scenario presented here, the current limits on $T^{0\nu\beta\beta}_{1/2}$ already constrain $\left| C_{ee} \right|$ to be smaller than about $10^{-3}$). At the same time, a detection of $0\nu\beta\beta$ decay in an ongoing or a future experiment, combined with an independent measurement excluding the IO for neutrino masses from neutrino oscillation experiments, would suggest a short-distance physics origin for the $0\nu\beta\beta$ decay signal.  Very large $0\nu\beta\beta$ signal would also not be possible from light-neutrino exchange, as it requires a degenerate neutrino mass spectrum with $m>0.1$ eV, which is clearly disfavored by present cosmological data.

Another consequence of the operator $\mathcal{O}^9$ leading to $0\nu\beta\beta$ decays is that the two emitted electrons would be right handed (as opposed to the usual contribution from light-neutrino exchange, where the two emitted electrons are left handed). A (hypothetical) measurement of the chirality of the emitted electrons would allow us to further test $\mathcal{O}^9$ as responsible for LNV and  generation of Majorana neutrino masses.

%%%%%%%%%%%%%%%%%%%%%%%%%%%%%%%%%%%%%%%%%%%%%%%%%%%%%%%%%%%%
\section{Conclusions} \label{sec:conclusion}
%%%%%%%%%%%%%%%%%%%%%%%%%%%%%%%%%%%%%%%%%%%%%%%%%%%%%%%%

Links between the origin of neutrino masses, the observed structure of neutrino masses and mixings, and $0\nu\beta\beta$ decay probes of the Majorana nature of neutrinos may provide a key to a unified understanding of these various central aspects of neutrino physics. Potentially, all these concepts are linked and generated by the same underlying new physics.  We have explored a simple class of theories beyond the SM that explain the large existing hierarchy between  the scale of neutrino masses and the electroweak scale by radiatively induced neutrino masses (exemplified by explicit known 2-loop and 3-loop scenarios) and that connect to specific neutrino mixing properties due to approximate zeroes in the neutrino mass matrix $m^{\nu}_{ab}$. In these theories, new physics beyond the SM responsible for LNV and the generation of Majorana neutrino masses does not couple directly  to quarks or left-handed leptons but couples to right-handed leptons. The consequence is that the leading (and dominant) 
contributions to lepton number violation and neutrino masses are encoded in a single nonrenormalizable dimension-9 operator, $\mathcal{O}^9$ in Eq.~(\ref{D9}). 

\smallskip

The present analysis shows the way approximate texture zeroes $m^{\nu}_{ee} \simeq 0$ and $m^{\nu}_{e\mu} \simeq 0$ naturally emerge in this class of theories. We have also defined a measure of the amount of hierarchy needed in the underlying  Yukawa matrix generating the neutrino mass matrix. Once $m^{\nu}_{ee} \simeq 0$ and $m^{\nu}_{e\mu} \simeq 0$ the rest of the entries of $m^{\nu}_{ab}$ are required to be of the same size in order to be compatible with current neutrino oscillation data, which then requires a mild amount of hierarchy in the Yukawa couplings (although less than the one present in the charged-lepton sector of the SM).

\smallskip

The texture with zeroes $m^{\nu}_{ee} \simeq 0$ and $m^{\nu}_{e\mu} \simeq 0$ generated in this class of theories gives rise to nontrivial correlations among the different neutrino parameters. These correlations can accommodate the current data from neutrino oscillation experiments while giving testable predictions for the unknowns in the neutrino sector, such as a normal neutrino mass ordering and a strong correlation among the values of the reactor angle $\theta_{13}$, the atmospheric angle $\theta_{23}$ (and its octant), and the $CP$ phase $\delta$ (see Figs.~\ref{fig1:ee}-\ref{fig3:texture} for various correlations). We also show that future, more precise measurements of $\left|\Delta m^2_{31}\right|$ and the value of the solar angle $\theta_{12}$ will be crucial for an ultimate test of this correlation.
         
\smallskip
         
Finally, this class of theories also incorporates an important link to $0\nu\beta\beta$ decay, since a generic feature of these scenarios is the existence of a contribution to $0\nu\beta\beta$ decay from short-distance physics which  largely dominates over the one coming from light-neutrino exchange (extremely suppressed in these scenarios, since $m^{\nu}_{ee} \simeq 0$) and can be sizable in many cases. We have analyzed the features of this leading, short-distance contribution to $0\nu\beta\beta$ decay in concrete scenarios and derived prospects of detection in both ongoing and upcoming $0\nu\beta\beta$ decay experiments such as GERDA, EXO, SNO+, KamLAND-Zen, CUORE, NEXT, MAJORANA, and SuperNEMO. Remarkably, these scenarios may  provide a detectable signal in $0\nu\beta\beta$ decay together with a normal ordering for neutrino masses and a lightest neutrino mass $m \lesssim 0.01$ eV. A future combination of data from neutrino oscillation experiments, cosmology, and $0\nu\beta\beta$ decay could allow us to 
ultimately test $\mathcal{O}^9$ as responsible for lepton number vialotaion and the generation of neutrino masses.

\vspace{7mm}

\begin{center}
\textbf{Acknowledgements} 
\end{center}

We thank M.\ Maltoni and T.\ Schwetz for sharing with us data from their global fits \cite{GonzalezGarcia:2012sz}. M.G. thanks T.~Schwetz for useful discussions and comments. J.M.N.\ thanks S.\ Pascoli for useful discussions, and A.\ Merle for discussions and useful comments on the manuscript. M.G.\ is supported by the Belgian Science Policy (IAP VI/11), the IISN, and the ARC project `Beyond Einstein: fundamental aspects of gravitational interactions'. J.M.N.\ is supported by the Science Technology and Facilities Council (STFC) under grant No. ST/J000477/1. M.R.\ is supported by Fondecyt grant No. 11110472, Anillo "Atlas Andino" ACT1102, and DGIP grant No. 11.12.39. 

\appendix
%%%%%%%%%%%%%%%%%%%%%%%%%%%%%%%%%%%%%%%%%%%%%%%%%%%%%%%%%%%%
\section{Master formula for $m_{ee}=m_{e\mu} =0$}\label{sec:app1}
\setcounter{equation}{0}
\renewcommand{\theequation}{A.\arabic{equation}}
%%%%%%%%%%%%%%%%%%%%%%%%%%%%%%%%%%%%%%%%%%%%%%%%%%%%%%%%%%%%

Equations (2.5) and (2.6), from the texture $m_{ee}=m_{e\mu} =0$,  can be rewritten as 
\bea
\frac{m_1}{m_3}   &=&   - e^{2 i \alpha_2}  \frac{s_{13}}{c_{13}^2} \left(s_{13} - t_{12} 
t_{23}  e^{i \delta}\right) \equiv A \label{eq:m1m3} \\
\frac{m_2}{m_3}   &=&  - e^{2 i (\alpha_2- \alpha_1)} 
\frac{s_{13}}{c_{13}^2} \left(s_{13} + t_{12}^{-1} t_{23}  e^{i \delta} \right) \equiv B.  \label{eq:m2m3}
\eea
Equations (\ref{eq:m1m3})  and  (\ref{eq:m2m3}) can then be expressed as
\bea
&|m_1|^2 &= \;   \frac{\Delta m^2_{31} |A|^2}{1 - |A|^2} \;\;\text{or}\;\; |m_3|^2 = \;   
\frac{\Delta m^2_{31}}{1-|A|^2} , 
 \label{eq:m} \\
&\cos{\delta} &=\; 
-\frac{ t_{23}^2 (1 - t_{12}^4) + t_{12}^2 (1 + t_{23}^2 t_{12}^2 - t_{13}^{-2}) \frac{\Delta m^2_{21}}{\Delta m^2_{31}}  }
{2 s_{13} t_{12} t_{23} ( 1 +  t_{12}^2  - t_{12}^2 \frac{\Delta m^2_{21}}{\Delta m^2_{31}} )}
\label{eq:cd}
\eea
where 

\be
|A|^2 = 
\frac{s_{13}^2 (s_{13}^2 + t_{12}^2 t_{23}^2  - 2  s_{13} t_{12} t_{23} \cos{\delta})}{c_{13}^4}
\ee

\begin{figure}[ht!]
\center{
\includegraphics[width=.32 \columnwidth]{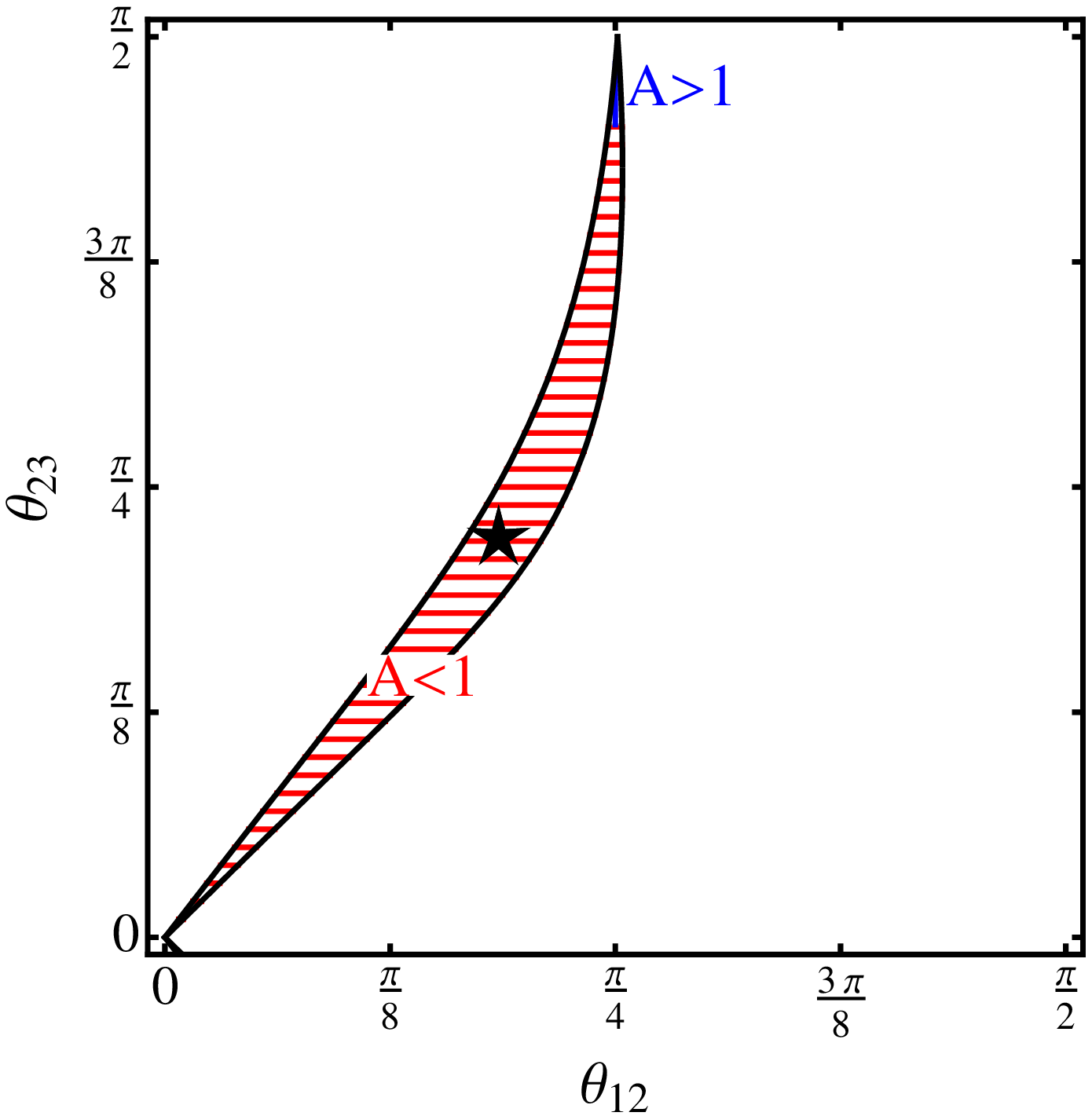}
\includegraphics[width=.32 \columnwidth]{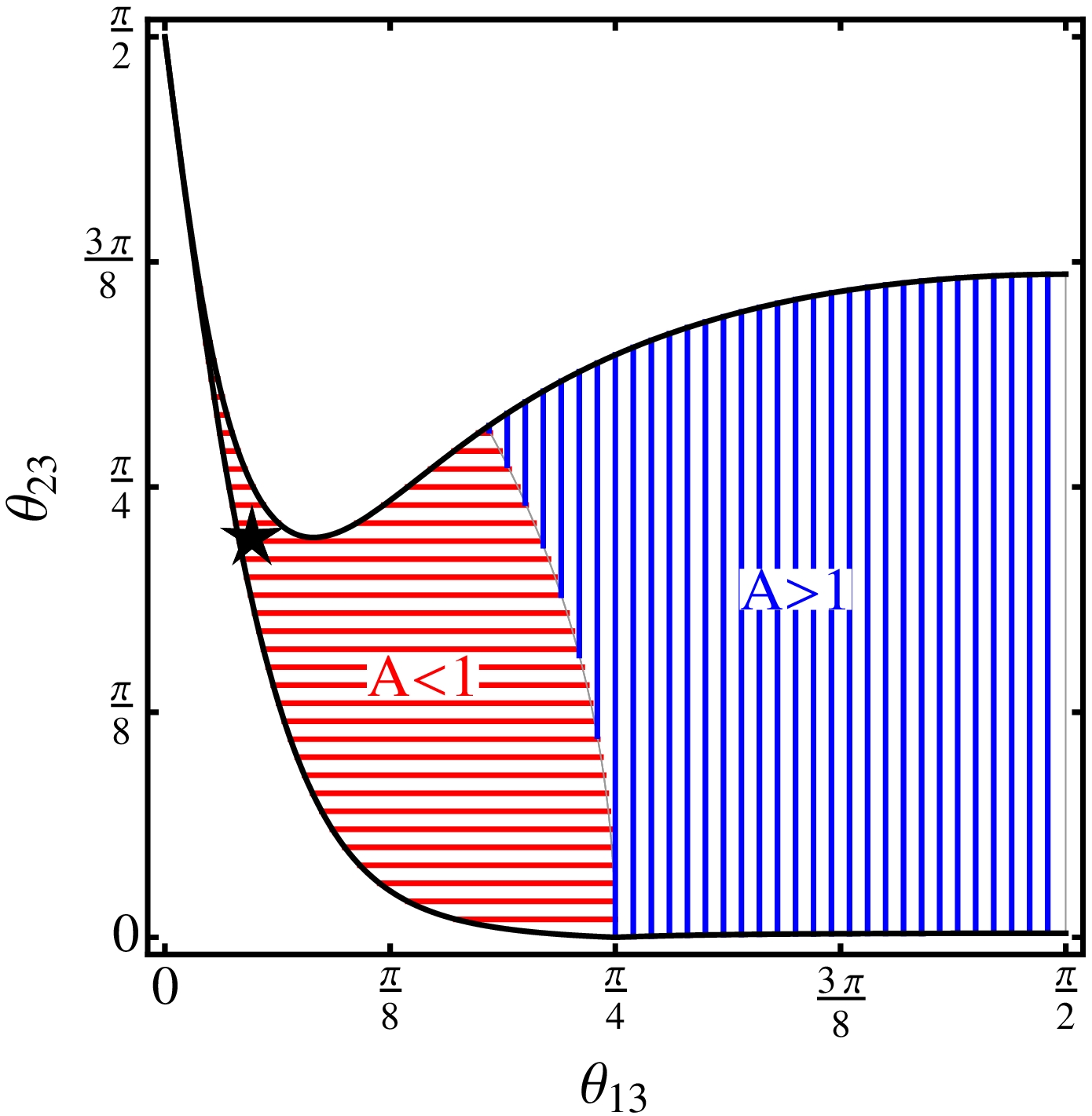}
\includegraphics[width=.33 \columnwidth]{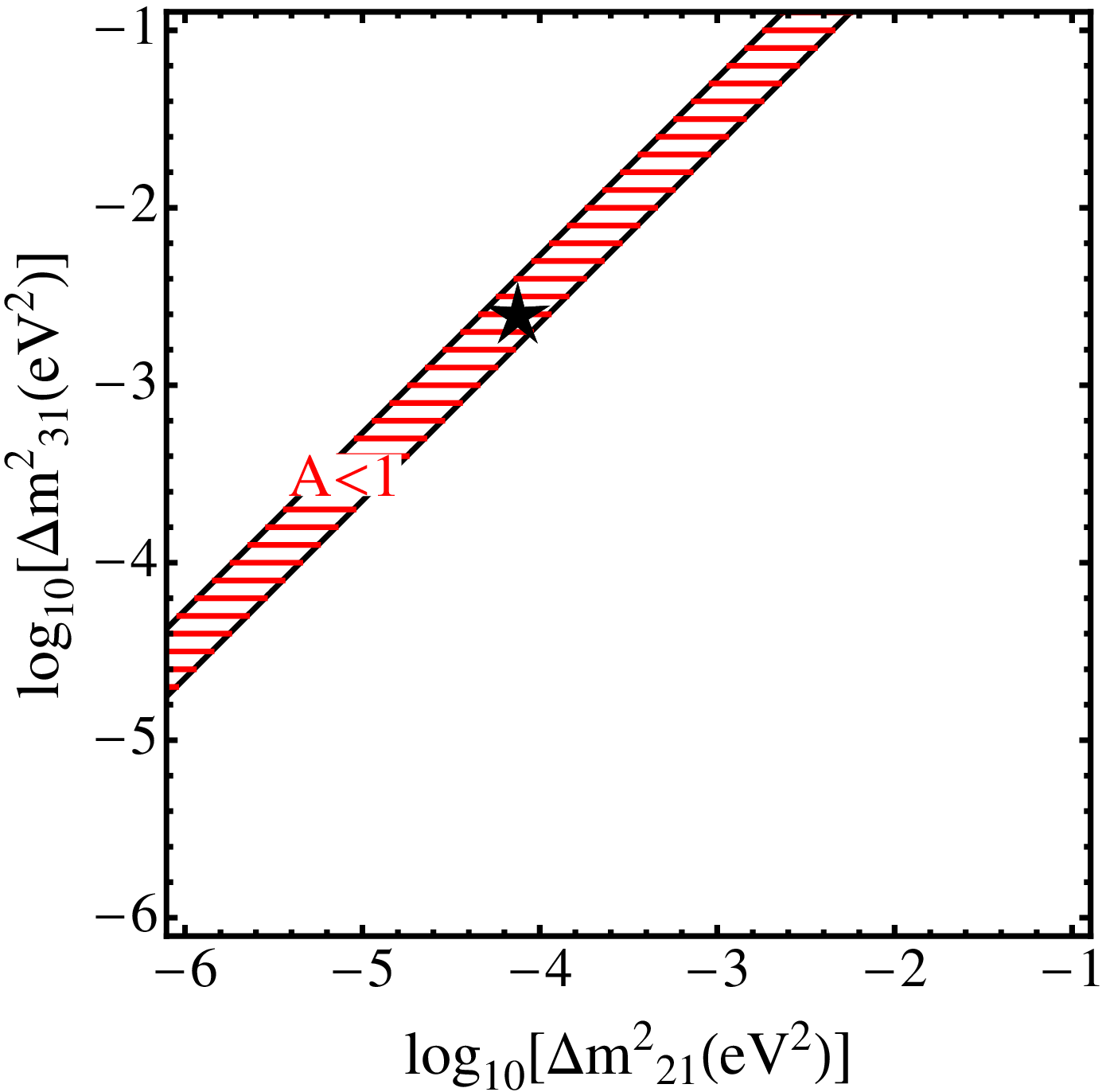}
}
\caption{\small 
For $m_{ee}=m_{e\mu} =0$, the allowed region $|\!\cos{\delta}|  \leq 1$ from Eq.~(\ref{eq:cd}). LEFT: in the ($\theta_{12}$, $\theta_{23}$) plane. MIDDLE: in the ($\theta_{13}$, $\theta_{23}$) plane. RIGHT: in the ($\Delta m^2_{21}$, $\Delta m^2_{31}$) plane. All slices are taken at the point of the remaining neutrino oscillation parameters fixed to their best-fit values from Ref.~\cite{GonzalezGarcia:2012sz}. Normal ordering ($|A| < 1$) and inverted ordering ($|A| >1$) are shown in the plots as (red) horizontal-hashed  and (blue) vertical-hashed regions, respectively. The black star corresponds to the experimentally  best-fit value of the neutrino oscillation parameters \cite{GonzalezGarcia:2012sz}.\label{fig:appendix}}
\end{figure}

A convenient rewriting of Eq.~(\ref{eq:cd}) then gives (\ref{master_1}). If Eqs.~(\ref{eq:m}) and (\ref{eq:cd}) are fulfilled, then Eqs.~(\ref{eq:m1m3}) and (\ref{eq:m2m3}) can always be satisfied with proper chosen values of $\alpha_{1,2}$:   
\bea
\tan{2 \alpha_1}                  &=&  \frac{-\sin{\delta}}{\cos{\delta} - s_{13} t_{12}^{-1} t_{23}^{-1}} \\
\tan{2 (\alpha_2-\alpha_1)} &=& \frac{-\sin{\delta}}{\cos{\delta} + s_{13} t_{12} t_{23}^{-1}}
\eea
Since $\alpha_{1}$ and $\alpha_{1}\!+\!\alpha_{2}$ are still free up to a multiple of $\pi/2$, this allows us to also choose the sign of $A$ and $B$. The allowed region of neutrino parameters can now comprehensively be expressed as
\bea
|\!\cos{\delta}|  \leq 1 \quad \text{with} \quad
\left\lbrace
\begin{array}{@{} l c @{}}
 |A| < 1 \; \Leftrightarrow \; \mathrm{(NO)}   \vspace{2mm}  \\ 
 |A| > 1 \; \Leftrightarrow \;\mathrm{(IO)}    
\end{array}
\right.\label{eq:master}
\eea
Fig.~\ref{fig:appendix} shows two-dimensional slices of the region in Eq.~(\ref{eq:master}). The slices are taken at the experimental best-fit point  of the neutrino oscillation parameters %(for those not shown in each slice) 
given in Ref.~\cite{GonzalezGarcia:2012sz}. The figure illustrates the nontrivial statement that a neutrino mass matrix texture with $m_{ee}=m_{e\mu} =0$ would be compatible with the neutrino oscillation data.

%%%%%%%%%%%%%%%%
%%%%%%%%%%%%%%%%


\begin{thebibliography}{99}
\bibitem{Mohapatra:2005wg}
  R.~N.~Mohapatra {\it et al.},
  %``Theory of neutrinos: A White paper,''
  Rept.\ Prog.\ Phys.\  {\bf 70} (2007) 1757
  [hep-ph/0510213].

\bibitem{Zee:1980ai}
  A.~Zee,
  %``A Theory of Lepton Number Violation, Neutrino Majorana Mass, and Oscillation,''
  Phys.\ Lett.\ B {\bf 93} (1980) 389
   [Erratum-ibid.\ B {\bf 95} (1980) 461]; Nucl.\ Phys.\ B {\bf 264} (1986) 99.
   
\bibitem{Babu:1988ki}
  K.~S.~Babu,
  %``Model of 'Calculable' Majorana Neutrino Masses,''
  Phys.\ Lett.\ B {\bf 203} (1988) 132.
  
 \bibitem{Krauss:2002px}
 L.~M.~Krauss, S.~Nasri and M.~Trodden,
 %``A Model for neutrino masses and dark matter,''
 Phys.\ Rev.\ D {\bf 67} (2003) 085002
 [hep-ph/0210389].
 
 \bibitem{Ma:2006km}
  E.~Ma,
  %``Verifiable radiative seesaw mechanism of neutrino mass and dark matter,''
  Phys.\ Rev.\ D {\bf 73} (2006) 077301
  [hep-ph/0601225].
  
  \bibitem{Aoki:2008av}
  M.~Aoki, S.~Kanemura and O.~Seto,
  %``Neutrino mass, Dark Matter and Baryon Asymmetry via TeV-Scale Physics without Fine-Tuning,''
  Phys.\ Rev.\ Lett.\  {\bf 102} (2009) 051805
  [arXiv:0807.0361 [hep-ph]].
  
\bibitem{Morisi:2012fg}
  S.~Morisi and J.~W.~F.~Valle,
  %``Neutrino masses and mixing: a flavour symmetry roadmap,''
  Fortsch.\ Phys.\  {\bf 61} (2013) 466
  [arXiv:1206.6678 [hep-ph]].  
  
 \bibitem{Hall:1999sn}
  L.~J.~Hall, H.~Murayama and N.~Weiner,
  %``Neutrino mass anarchy,''
  Phys.\ Rev.\ Lett.\  {\bf 84} (2000) 2572
  [hep-ph/9911341].

\bibitem{Fritzsch:1977za} 
  H.~Fritzsch,
  %``Calculating the Cabibbo Angle,''
  Phys.\ Lett.\ B {\bf 70}, 436 (1977).
\bibitem{Fritzsch:1977vd} 
  H.~Fritzsch,
  %``Weak Interaction Mixing in the Six - Quark Theory,''
  Phys.\ Lett.\ B {\bf 73}, 317 (1978).
\bibitem{Weinberg:1977hb} 
  S.~Weinberg,
  %``The Problem of Mass,''
  Trans.\ New York Acad.\ Sci.\  {\bf 38}, 185 (1977).
\bibitem{Wilczek:1977uh} 
  F.~Wilczek and A.~Zee,
  %``Discrete Flavor Symmetries and a Formula for the Cabibbo Angle,''
  Phys.\ Lett.\ B {\bf 70}, 418 (1977)
  [Erratum-ibid.\  {\bf 72B}, 504 (1978)].

\bibitem{Frampton:2002yf}
  P.~H.~Frampton, S.~L.~Glashow and D.~Marfatia,
  %``Zeroes of the neutrino mass matrix,''
  Phys.\ Lett.\ B {\bf 536} (2002) 79
  [hep-ph/0201008].

\bibitem{Xing:2002ta} 
  Z.~-z.~Xing,
  %``Texture zeros and Majorana phases of the neutrino mass matrix,''
  Phys.\ Lett.\ B {\bf 530}, 159 (2002)
  [hep-ph/0201151].
\bibitem{Xing:2002ap} 
  Z.~-z.~Xing,
  %``A Full determination of the neutrino mass spectrum from two zero textures of the neutrino mass matrix,''
  Phys.\ Lett.\ B {\bf 539}, 85 (2002)
  [hep-ph/0205032].
 \bibitem{Desai:2002sz} 
  B.~R.~Desai, D.~P.~Roy and A.~R.~Vaucher,
  %``Three neutrino mass matrices with two texture zeros,''
  Mod.\ Phys.\ Lett.\ A {\bf 18}, 1355 (2003)
  [hep-ph/0209035].
\bibitem{Guo:2002ei}
  W.~-l.~Guo and Z.~-z.~Xing,
  %``Implications of the KamLAND measurement on the lepton flavor mixing matrix and the neutrino mass matrix,''
  Phys.\ Rev.\ D {\bf 67} (2003) 053002
  [hep-ph/0212142].  
\bibitem{Honda:2003pg} % Check stability of th13 > prediction 
  M.~Honda, S.~Kaneko and M.~Tanimoto,
  %``Prediction and its stability in neutrino mass matrix with two zeros,''
  JHEP {\bf 0309}, 028 (2003)
  [hep-ph/0303227].  
\bibitem{Merle:2006du} %updated plots
  A.~Merle and W.~Rodejohann,
  %``The Elements of the neutrino mass matrix: Allowed ranges and implications of texture zeros,''
  Phys.\ Rev.\ D {\bf 73}, 073012 (2006)
  [hep-ph/0603111].
\bibitem{Dev:2006xu} 
  S.~Dev, S.~Kumar, S.~Verma and S.~Gupta,
  %``Phenomenological implications of a class of neutrino mass matrices,''
  Nucl.\ Phys.\ B {\bf 784}, 103 (2007)
  [hep-ph/0611313].  
\bibitem{Dev:2006qe} 
  S.~Dev, S.~Kumar, S.~Verma and S.~Gupta,
  %``Phenomenology of two-texture zero neutrino mass matrices,''
  Phys.\ Rev.\ D {\bf 76}, 013002 (2007)
  [hep-ph/0612102].  
\bibitem{Fritzsch:2011qv} 
  H.~Fritzsch, Z.~-z.~Xing and S.~Zhou,
  %``Two-zero Textures of the Majorana Neutrino Mass Matrix and Current Experimental Tests,''
  JHEP {\bf 1109}, 083 (2011)
  [arXiv:1108.4534 [hep-ph]].
\bibitem{Kumar:2011vf} 
  S.~Kumar,
  %``Implications of a class of neutrino mass matrices with texture zeros for non-zero $\theta_{13}$,''
  Phys.\ Rev.\ D {\bf 84}, 077301 (2011)
  [arXiv:1108.2137 [hep-ph]].
\bibitem{Ludl:2011vv}
  P.~O.~Ludl, S.~Morisi and E.~Peinado,
  %``The Reactor mixing angle and CP violation with two texture zeros in the light of T2K,''
  Nucl.\ Phys.\ B {\bf 857} (2012) 411
  [arXiv:1109.3393 [hep-ph]].  
\bibitem{Meloni:2012sx} 
  G.~Blankenburg  and D.~Meloni,
  %``Fine-tuning and naturalness issues in the two-zero neutrino mass textures,''
  Nucl.\ Phys.\ B {\bf 867}, 749 (2013)
  [arXiv:1204.2706 [hep-ph]].
\bibitem{Grimus:2012zm} %Mass matrix
  W.~Grimus and P.~O.~Ludl,
  %``Two-parameter neutrino mass matrices with two texture zeros,''
  J.\ Phys.\ G {\bf 40} (2013) 055003
  [arXiv:1208.4515 [hep-ph]].
\bibitem{Grimus:2012ii} % upper and lower limits on m_ab
  W.~Grimus and P.~O.~Ludl,
  %``Correlations of the elements of the neutrino mass matrix,''
  JHEP {\bf 1212} (2012) 117
  [arXiv:1209.2601 [hep-ph]].


\cite{Meloni:2012sx,Grimus:2012zm}
%\bibitem{Kageyama:2002zw} %% See-saw realisation
%  A.~Kageyama, S.~Kaneko, N.~Shimoyama and M.~Tanimoto,
%  %``Seesaw realization of the texture zeros in the neutrino mass matrix,''
%  Phys.\ Lett.\ B {\bf 538} (2002) 96
%  [hep-ph/0204291].  
%\bibitem{Grimus:2004hf} % Abbelian gauge symmtry  to texture zeros 
%  W.~Grimus, A.~S.~Joshipura, L.~Lavoura and M.~Tanimoto,
%  %``Symmetry realization of texture zeros,''
%  Eur.\ Phys.\ J.\ C {\bf 36}, 227 (2004)
%  [hep-ph/0405016].  


\bibitem{Chen:2006vn}
  C.~-S.~Chen, C.~Q.~Geng and J.~N.~Ng,
  %``Unconventional Neutrino Mass Generation, Neutrinoless Double Beta Decays, and Collider Phenomenology,''
  Phys.\ Rev.\ D {\bf 75} (2007) 053004
  [hep-ph/0610118].

\bibitem{delAguila:2011gr}
  F.~del Aguila, A.~Aparici, S.~Bhattacharya, A.~Santamaria and J.~Wudka,
  %``A realistic model of neutrino masses with a large neutrinoless double beta decay rate,''
  JHEP {\bf 1205} (2012) 133
  [arXiv:1111.6960 [hep-ph]].

\bibitem{delAguila:2012nu}
  F.~del Aguila, A.~Aparici, S.~Bhattacharya, A.~Santamaria and J.~Wudka,
  %``Effective Lagrangian approach to neutrinoless double beta decay and neutrino masses,''
  JHEP {\bf 1206} (2012) 146
  [arXiv:1204.5986 [hep-ph]].

\bibitem{Gustafsson:2012vj}
  M.~Gustafsson, J.~M.~No and M.~A.~Rivera,
  %``The Cocktail Model: Neutrino Masses and Mixings with Dark Matter,''
  Phys.\ Rev.\ Lett.\  {\bf 110} (2013) 211802
  [arXiv:1212.4806 [hep-ph]].

\bibitem{Farzan:2012ev}
  Y.~Farzan, S.~Pascoli and M.~A.~Schmidt,
  %``Recipes and Ingredients for Neutrino Mass at Loop Level,''
  JHEP {\bf 1303} (2013) 107
  [arXiv:1208.2732 [hep-ph]].
  
\bibitem{Schonert:2005zn}
  S.~.Schonert {\it et al.}  [GERDA Collaboration],
  %``The GERmanium Detector Array (GERDA) for the search of neutrinoless beta beta decays of Ge-76 at LNGS,''
  Nucl.\ Phys.\ Proc.\ Suppl.\  {\bf 145} (2005) 242.


\bibitem{Agostini:2013mzu}
  M.~Agostini {\it et al.}  [GERDA Collaboration],
  %``Results on neutrinoless double beta decay of 76Ge from GERDA Phase I,''
  Phys.\ Rev.\ Lett.\  {\bf 111} (2013) 122503
  [arXiv:1307.4720 [nucl-ex]].

\bibitem{Akimov:2005mq}
  D.~Akimov {\it et al.},
  %``EXO: An advanced Enriched Xenon double-beta decay Observatory,''
  Nucl.\ Phys.\ Proc.\ Suppl.\  {\bf 138} (2005) 224.

\bibitem{Chen:2005yi}
  M.~C.~Chen,
  %``The SNO liquid scintillator project,''
  Nucl.\ Phys.\ Proc.\ Suppl.\  {\bf 145} (2005) 65.

 \bibitem{Mitsui:2011zza}
  T.~Mitsui [KamLAND Collaboration],
  %``Low-energy neutrino physics with KamLAND,''
  Nucl.\ Phys.\ Proc.\ Suppl.\  {\bf 217} (2011) 89.
  
  \bibitem{KamLANDZen:2012aa}
  A.~Gando {\it et al.}  [KamLAND-Zen Collaboration],
  %``Measurement of the double-\beta decay half-life of ^{136}Xe with the KamLAND-Zen experiment,''
  Phys.\ Rev.\ C {\bf 85} (2012) 045504
  [arXiv:1201.4664 [hep-ex]].

\bibitem{Arnaboldi:2002du}
  C.~Arnaboldi {\it et al.}  [CUORE Collaboration],
  %``CUORE: A Cryogenic underground observatory for rare events,''
  Nucl.\ Instrum.\ Meth.\ A {\bf 518} (2004) 775
  [hep-ex/0212053];
  R.~Ardito {\it et al.},
  %``CUORE: A Cryogenic underground observatory for rare events,''
  hep-ex/0501010.

\bibitem{Dafni:2012nwa}
F.~Granena {\it et al.}  [NEXT Collaboration],
  %``NEXT, a HPGXe TPC for neutrinoless double beta decay searches,''
  arXiv:0907.4054 [hep-ex].
  
  \bibitem{GomezCadenas:2012jv}
  J.~J.~Gomez-Cadenas, J.~Martin-Albo and F.~Monrabal,
  %``NEXT, high-pressure xenon gas experiments for ultimate sensitivity to Majorana neutrinos,''
  JINST {\bf 7} (2012) C11007
  [arXiv:1210.0341 [physics.ins-det]].

\bibitem{Gaitskell:2003zr}
  R.~Gaitskell {\it et al.}  [Majorana Collaboration],
  %``The Majorana zero neutrino double beta decay experiment,''
  nucl-ex/0311013.
  
\bibitem{Arnold:2010tu}
  R.~Arnold {\it et al.}  [SuperNEMO Collaboration],
  %``Probing New Physics Models of Neutrinoless Double Beta Decay with SuperNEMO,''
  Eur.\ Phys.\ J.\ C {\bf 70} (2010) 927
  [arXiv:1005.1241 [hep-ex]].

\bibitem{Pontecorvo:1957qd}
 B.~Pontecorvo,
%``Inverse beta processes and nonconservation of lepton charge,''
Sov.\ Phys.\ JETP {\bf 7} (1958) 172
   [Zh.\ Eksp.\ Teor.\ Fiz.\  {\bf 34} (1957) 247];
 Z.~Maki, M.~Nakagawa and S.~Sakata,
  %``Remarks on the unified model of elementary particles,''
  Prog.\ Theor.\ Phys.\  {\bf 28} (1962) 870.

\bibitem{Harrison:2002er}
  P.~F.~Harrison, D.~H.~Perkins and W.~G.~Scott,
  %``Tri-bimaximal mixing and the neutrino oscillation data,''
  Phys.\ Lett.\ B {\bf 530} (2002) 167
  [hep-ph/0202074].

\bibitem{An:2012eh}
  F.~P.~An {\it et al.}  [DAYA-BAY Collaboration],
  %``Observation of electron-antineutrino disappearance at Daya Bay,''
  Phys.\ Rev.\ Lett.\  {\bf 108} (2012) 171803
  [arXiv:1203.1669 [hep-ex]].

\bibitem{Ahn:2012nd}
  J.~K.~Ahn {\it et al.}  [RENO Collaboration],
  %``Observation of Reactor Electron Antineutrino Disappearance in the RENO Experiment,''
  Phys.\ Rev.\ Lett.\  {\bf 108} (2012) 191802
  [arXiv:1204.0626 [hep-ex]].

\bibitem{Abe:2011fz}
  Y.~Abe {\it et al.}  [DOUBLE-CHOOZ Collaboration],
  %``Indication for the disappearance of reactor electron antineutrinos in the Double Chooz experiment,''
  Phys.\ Rev.\ Lett.\  {\bf 108} (2012) 131801
  [arXiv:1112.6353 [hep-ex]];
Y.~Abe {\it et al.}  [Double Chooz Collaboration],
  %``Reactor electron antineutrino disappearance in the Double Chooz experiment,''
  Phys.\ Rev.\ D {\bf 86} (2012) 052008
  [arXiv:1207.6632 [hep-ex]].

\bibitem{Abe:2011sj}
  K.~Abe {\it et al.}  [T2K Collaboration],
  %``Indication of Electron Neutrino Appearance from an Accelerator-produced Off-axis Muon Neutrino Beam,''
  Phys.\ Rev.\ Lett.\  {\bf 107} (2011) 041801
  [arXiv:1106.2822 [hep-ex]].

\bibitem{Adamson:2011qu}
  P.~Adamson {\it et al.}  [MINOS Collaboration],
  %``Improved search for muon-neutrino to electron-neutrino oscillations in MINOS,''
  Phys.\ Rev.\ Lett.\  {\bf 107} (2011) 181802
  [arXiv:1108.0015 [hep-ex]];
 P.~Adamson {\it et al.}  [MINOS Collaboration],
  %``An improved measurement of muon antineutrino disappearance in MINOS,''
  Phys.\ Rev.\ Lett.\  {\bf 108} (2012) 191801
  [arXiv:1202.2772 [hep-ex]].

\bibitem{Wendell:2010md}
  R.~Wendell {\it et al.}  [Super-Kamiokande Collaboration],
  %``Atmospheric neutrino oscillation analysis with sub-leading effects in Super-Kamiokande I, II, and III,''
  Phys.\ Rev.\ D {\bf 81} (2010) 092004
  [arXiv:1002.3471 [hep-ex]].

\bibitem{Itow:2013zza}
  Y.~Itow,
  %``Recent results in atmospheric neutrino oscillations in the light of large $\Theta_{13}$,''
  Nucl.\ Phys.\ Proc.\ Suppl.\  {\bf 235-236} (2013) 79.

\bibitem{Tortola:2012te}
  D.~V.~Forero, M.~Tortola and J.~W.~F.~Valle,
  %``Global status of neutrino oscillation parameters after Neutrino-2012,''
  Phys.\ Rev.\ D {\bf 86} (2012) 073012
  [arXiv:1205.4018 [hep-ph]].

\bibitem{Fogli:2012ua}
  G.~L.~Fogli, E.~Lisi, A.~Marrone, D.~Montanino, A.~Palazzo and A.~M.~Rotunno,
  %``Global analysis of neutrino masses, mixings and phases: entering the era of leptonic CP violation searches,''
  Phys.\ Rev.\ D {\bf 86} (2012) 013012
  [arXiv:1205.5254 [hep-ph]].

\bibitem{GonzalezGarcia:2012sz}
  M.~C.~Gonzalez-Garcia, M.~Maltoni, J.~Salvado and T.~Schwetz,
  %``Global fit to three neutrino mixing: critical look at present precision,''
  JHEP {\bf 1212} (2012) 123
  [arXiv:1209.3023 [hep-ph]].

\bibitem{Ade:2013zuv}
  P.~A.~R.~Ade {\it et al.}  [Planck Collaboration],
  %``Planck 2013 results. XVI. Cosmological parameters,''
  arXiv:1303.5076 [astro-ph.CO].

\bibitem{Pascoli:2001by}
  S.~Pascoli, S.~T.~Petcov and L.~Wolfenstein,
  %``Searching for the CP violation associated with Majorana neutrinos,''
  Phys.\ Lett.\ B {\bf 524} (2002) 319
  [hep-ph/0110287]; Z.~-z.~Xing,
  %``Vanishing effective mass of the neutrinoless double beta decay?,''
  Phys.\ Rev.\ D {\bf 68} (2003) 053002
  [hep-ph/0305195]; S.~Choubey and W.~Rodejohann,
  %``Neutrinoless double beta decay and future neutrino oscillation precision experiments,''
  Phys.\ Rev.\ D {\bf 72} (2005) 033016
  [hep-ph/0506102]; M.~Lindner, A.~Merle and W.~Rodejohann,
  %``Improved limit on theta(13) and implications for neutrino masses in neutrino-less double beta decay and cosmology,''
  Phys.\ Rev.\ D {\bf 73} (2006) 053005
  [hep-ph/0512143].

\bibitem{NUFIT}
http://www.nu-fit.org/?q=node/8

\bibitem{Babu:2001ex}
  K.~S.~Babu and C.~N.~Leung,
  %``Classification of effective neutrino mass operators,''
  Nucl.\ Phys.\ B {\bf 619} (2001) 667
  [hep-ph/0106054].

\bibitem{Gouvea:2007xp}
  A.~de Gouvea and J.~Jenkins,
  %``A Survey of Lepton Number Violation Via Effective Operators,''
  Phys.\ Rev.\ D {\bf 77} (2008) 013008
  [arXiv:0708.1344 [hep-ph]].

\bibitem{Angel:2012ug}
  P.~W.~Angel, N.~L.~Rodd and R.~R.~Volkas,
  %``Origin of neutrino masses at the LHC: $\Delta L = 2$ effective operators and their ultraviolet completions,''
  Phys.\ Rev.\ D {\bf 87} (2013) 7,  073007
  [arXiv:1212.6111 [hep-ph]].

\bibitem{Weinberg:1979sa}
  S.~Weinberg,
  %``Baryon and Lepton Nonconserving Processes,''
  Phys.\ Rev.\ Lett.\  {\bf 43} (1979) 1566.
  
\bibitem{Chang:1999hga} 
  D.~Chang and A.~Zee,
  %``Radiatively induced neutrino Majorana masses and oscillation,''
  Phys.\ Rev.\ D {\bf 61}, 071303 (2000)
  [hep-ph/9912380].  
  

\bibitem{MATLAB}
MATLAB version 7.12.0.635 (R2011a) , The MathWorks, Inc., Natick, Massachusetts, United States.
  
\bibitem{Aparici:2013xga}
  A.~Aparici,
  %``Exotic properties of neutrinos using effective Lagrangians and specific models,''
  arXiv:1312.0554 [hep-ph].
 
  
\bibitem{Chen:2007dc}
  C.~-S.~Chen, C.~-Q.~Geng, J.~N.~Ng and J.~M.~S.~Wu,
  %``Testing radiative neutrino mass generation at the LHC,''
  JHEP {\bf 0708} (2007) 022
  [arXiv:0706.1964 [hep-ph]].

  \bibitem{Bellgardt:1987du}
  U.~Bellgardt {\it et al.}  [SINDRUM Collaboration],
  %``Search for the Decay mu+ ---> e+ e+ e-,''
  Nucl.\ Phys.\ B {\bf 299} (1988) 1.
  
\bibitem{Adam:2013mnn}
  J.~Adam {\it et al.}  [MEG Collaboration],
  %``New constraint on the existence of the mu+-> e+ gamma decay,''
  Phys.\ Rev.\ Lett.\  {\bf 110} (2013) 201801
  [arXiv:1303.0754 [hep-ex]]. 
  
\bibitem{Hayasaka:2010np}
  K.~Hayasaka {\it et al.},
  %``Search for Lepton Flavor Violating Tau Decays into Three Leptons with 719 Million Produced Tau+Tau- Pairs,''
  Phys.\ Lett.\ B {\bf 687} (2010) 139
  [arXiv:1001.3221 [hep-ex]]; J.~P.~Lees {\it et al.}  [BaBar Collaboration],
  %``Limits on tau Lepton-Flavor Violating Decays in three charged leptons,''
  Phys.\ Rev.\ D {\bf 81} (2010) 111101
  [arXiv:1002.4550 [hep-ex]]. 
    
\bibitem{Nebot:2007bc}
  M.~Nebot, J.~F.~Oliver, D.~Palao and A.~Santamaria,
  %``Prospects for the Zee-Babu Model at the CERN LHC and low energy experiments,''
  Phys.\ Rev.\ D {\bf 77} (2008) 093013
  [arXiv:0711.0483 [hep-ph]].  
  
\bibitem{Blennow:2010th}
  M.~Blennow, E.~Fernandez-Martinez, J.~Lopez-Pavon and J.~Menendez,
  %``Neutrinoless double beta decay in seesaw models,''
  JHEP {\bf 1007} (2010) 096
  [arXiv:1005.3240 [hep-ph]].
  
  
\bibitem{LopezPavon:2012zg} 
  J.~Lopez-Pavon, S.~Pascoli and C.~-f.~Wong,
  %``Can heavy neutrinos dominate neutrinoless double beta decay?,''
  Phys.\ Rev.\ D {\bf 87}, no. 9, 093007 (2013)
  [arXiv:1209.5342 [hep-ph]].  
  
 \bibitem{Bonnet:2012kh} 
  F.~Bonnet, M.~Hirsch, T.~Ota and W.~Winter,
  %``Systematic decomposition of the neutrinoless double beta decay operator,''
  JHEP {\bf 1303}, 055 (2013)
  [arXiv:1212.3045]. 

\bibitem{Pas:2000vn}
  H.~Pas, M.~Hirsch, H.~V.~Klapdor-Kleingrothaus and S.~G.~Kovalenko,
  %``A Superformula for neutrinoless double beta decay. 2. The Short range part,''
  Phys.\ Lett.\ B {\bf 498} (2001) 35
  [hep-ph/0008182].

\bibitem{Bergstrom:2011dt}
  J.~Bergstrom, A.~Merle and T.~Ohlsson,
  %``Constraining New Physics with a Positive or Negative Signal of Neutrino-less Double Beta Decay,''
  JHEP {\bf 1105} (2011) 122
  [arXiv:1103.3015 [hep-ph]].  
  
\bibitem{Suhonen:1998ck}
  J.~Suhonen and O.~Civitarese,
  %``Weak-interaction and nuclear-structure aspects of nuclear double beta decay,''
  Phys.\ Rept.\  {\bf 300} (1998) 123.
  
 \bibitem{GomezCadenas:2010gs}
  J.~J.~Gomez-Cadenas, J.~Martin-Albo, M.~Sorel, P.~Ferrario, F.~Monrabal, J.~Munoz-Vidal, P.~Novella and A.~Poves,
  %``Sense and sensitivity of double beta decay experiments,''
  JCAP {\bf 1106} (2011) 007
  [arXiv:1010.5112 [hep-ex]].  
  
 \bibitem{Deppisch:2012nb}
 F.~F.~Deppisch, M.~Hirsch and H.~Pas,
 %``Neutrinoless Double Beta Decay and Physics Beyond the Standard Model,''
 J.\ Phys.\ G {\bf 39} (2012) 124007
 [arXiv:1208.0727 [hep-ph]].
  
 \bibitem{Macolino:2013ifa}
  C.~Macolino [on behalf of the GERDA Collaboration],
  %``Results on neutrinoless double beta decay from GERDA Phase I,''
  Mod.\ Phys.\ Lett.\ A {\bf 29} (2014) 1430001
  [arXiv:1312.0562 [hep-ex]]. 
 
 \bibitem{Auger:2012ar}
  M.~Auger {\it et al.}  [EXO Collaboration],
  %``Search for Neutrinoless Double-Beta Decay in $^{136}$Xe with EXO-200,''
  Phys.\ Rev.\ Lett.\  {\bf 109} (2012) 032505
  [arXiv:1205.5608 [hep-ex]].

  \bibitem{Wamba:2005hr}
  K.~Wamba [EXO Collaboration],
  %``EXO: A Status Report,''
  SLAC-WP-068.
 
 \bibitem{Gando:2012zm}
  A.~Gando {\it et al.}  [KamLAND-Zen Collaboration],
  %``Limit on Neutrinoless $\beta\beta$ Decay of Xe-136 from the First Phase of KamLAND-Zen and Comparison with the Positive Claim in Ge-76,''
  Phys.\ Rev.\ Lett.\  {\bf 110} (2013) 6,  062502
  [arXiv:1211.3863 [hep-ex]].
  
  \bibitem{GomezCadenas:2011it}
  J.~J.~Gomez-Cadenas, J.~Martin-Albo, M.~Mezzetto, F.~Monrabal and M.~Sorel,
  %``The Search for neutrinoless double beta decay,''
  Riv.\ Nuovo Cim.\  {\bf 35} (2012) 29
  [arXiv:1109.5515 [hep-ex]].
  
  \bibitem{Argyriades:2008pr}
  J.~Argyriades {\it et al.}  [NEMO Collaboration],
  %``Measurement of the Double Beta Decay Half-life of Nd-150 and Search for Neutrinoless Decay Modes with the NEMO-3 Detector,''
  Phys.\ Rev.\ C {\bf 80} (2009) 032501
  [arXiv:0810.0248 [hep-ex]].
  
  \bibitem{Maneira:2013fsa}
  J.~éManeira [SNO+ Collaboration],
  %``The SNO$+$ experiment: status and overview,''
  J.\ Phys.\ Conf.\ Ser.\  {\bf 447} (2013) 012065.
  
  \bibitem{Arnaboldi:2008ds}
  C.~Arnaboldi {\it et al.}  [CUORICINO Collaboration],
  %``Results from a search for the 0 neutrino beta beta-decay of Te-130,''
  Phys.\ Rev.\ C {\bf 78} (2008) 035502
  [arXiv:0802.3439 [hep-ex]].
  
  \bibitem{Arnold:2005rz}
  R.~Arnold {\it et al.}  [NEMO Collaboration],
  %``First results of the search of neutrinoless double beta decay with the NEMO 3 detector,''
  Phys.\ Rev.\ Lett.\  {\bf 95} (2005) 182302
  [hep-ex/0507083].
  
  \bibitem{Barabash:2010bd}
  A.~S.~Barabash {\it et al.}  [NEMO Collaboration],
  %``Investigation of double beta decay with the NEMO-3 detector,''
  Phys.\ Atom.\ Nucl.\  {\bf 74} (2011) 312
  [arXiv:1002.2862 [nucl-ex]].
  
\end{thebibliography}
\end{document}